\documentclass[manuscript]{acmart}

\usepackage{xcolor} 

\definecolor{minor}{rgb}{0,0,0}
\definecolor{revision}{rgb}{0,0,0}
\definecolor{revision2}{rgb}{0,0,0}

\usepackage[normalem]{ulem}
\usepackage{longtable}
\usepackage{adjustbox}
\usepackage{pdfpages}

\AtBeginDocument{%
  \providecommand\BibTeX{{%
    \normalfont B\kern-0.5em{\scshape i\kern-0.25em b}\kern-0.8em\TeX}}}

\setcopyright{rightsretained} 
\acmJournal{TOCHI}
\acmYear{2022} \acmVolume{1} \acmNumber{1} \acmArticle{1} \acmMonth{1} \acmPrice{}\acmDOI{10.1145/3577012}

\begin{document}


\title{Seeing Through Things: Exploring the Design Space of Privacy-Aware Data-Enabled Objects}



\author{Yu-Ting Cheng}
\email{y.cheng@tue.nl}
\affiliation{
  \institution{Eindhoven University of Technology \& National Taiwan University of Science and Technology}
  \country{the Netherlands \& Taiwan}}

\author{Mathias Funk}
\affiliation{
  \institution{Eindhoven University of Technology}
  \country{the Netherlands}}
\email{m.funk@tue.nl}

\author{Rung-Huei Liang}
\email{liang@ntust.edu.tw}
\affiliation{%
  \institution{National Taiwan University of Science and Technology}
  \country{Taiwan}
}

\author{Lin-Lin Chen}
\email{l.chen@tue.nl}
\affiliation{%
 \institution{Eindhoven University of Technology}
 \country{the Netherlands}}

\renewcommand{\shortauthors}{Cheng, et al.}

\begin{abstract}
Increasing amounts of sensor-augmented research objects have been used in design research. We call these objects Data-Enabled Objects, which can be integrated into daily activities capturing data about people's detailed whereabouts, behaviours and routines. These objects provide data perspectives on everyday life for contextual design research. However, data-enabled objects are still computational devices with limited privacy awareness and nuanced data sharing. To better design data-enabled objects, we explore privacy design spaces by inviting 18 teams of undergraduate design students to re-design the same type of sensor-enabled home research camera. We developed the Connected Peekaboo Toolkit (CPT) to support the design teams in designing, building, and directly deploying their prototypes in real home studies. We conducted Thematic Analysis to analyse their outcomes which led us to interpret that privacy is not just an obstacle but can be a driver by unfolding an exploration of possible design spaces for data-enabled objects. 

\end{abstract}






\maketitle

\section{Introduction}
Ubiquitous sensing and massive data collection allow smart products, such as Amazon Echo, Google Home, or Apple HomePod, to quantify human behavior and thereby act intelligently by anticipating and responding to people's needs—at least, that is the promise. Parallel to these commercial developments, researchers have also developed sensor-augmented research objects such as Connected Resources~\cite{kitazaki_2019_CHIEA}, Connected Baby Bottle~\cite{Bogers_2016_DIS}, and Morse Things~\cite{wakkary_2017_DIS} to capture aspects of everyday activities in minute detail, shedding a new light on design ethnography. These sensor-augmented research objects are purposely-built instrumented objects that have finished product quality to be integrated into people's daily \textcolor{minor}{lives} to capture, store and share data from the field. In this paper, we call these sensor-augmented research objects~\textit{Data-Enabled Objects}\textcolor{minor}{, because they} capture data about people's natural interactions in the environment and with the objects. \textcolor{revision}{The need to collect data stems from the need to understand the everyday life context before and during designing for that context. For example, data can \textcolor{revision2}{be collected to} support design researchers in \textcolor{revision2}{studying people's unconscious or private behaviours (e.g., by using wearable devices to track sleep quality~\cite{nguyen_2020_nordichi} or alcohol consumption~\cite{you_2019_mobileHCI});}
in prompting people to have more grounded responses during interviews (e.g., by using data-enabled design~\cite{Bogers_2018_DIS, Bogers_2016_DIS} or ritual camera~\cite{mols_2016_IEEE}); and in engaging people to discover hidden practices in their familiar routines through more-than human-centred perspectives (e.g., by using thing ethnography~\cite{Giaccardi_2016_DIS, Chang_2017_DIS}, non-actor scenery photo~\cite{cheng_when_2021}, or thing constellation~\cite{huang_thing_2021}).}


Data-enabled objects are computational devices with on-board memory and processing capabilities. While they are capable of capturing long-term data consistently and remotely without fatigue, they can be insensitive to the context of that data and to changes taking place within that context. Unless explicitly programmed, they cannot adapt their data collection mechanisms to react to different situations. Data-enabled objects can act as a kind of black box that continuously and inconspicuously absorbs personal data, including bio-signals and other health-related data, detailed whereabouts, and personal preferences. At the same time, research participants may be lenient with research-purpose data collection because they may be acquainted with the researchers or have commitments to assist with the academic work that the research is part of. Thus, considerable damage can potentially result from any malpractice that takes place in the application of data-enabled objects, for example, through the leakage or misuse of such data, even unintentionally~\cite{Freed_2018_CHI, Pierce_2019_rtd, Pierce_2019_CHI}. Consequently, people involved in the research might experience anxiety contemplating the likelihood of such events~\cite{Pierce_2018_CHI}, rightfully attributed to the presence of the data-enabled objects in the home. Privacy issues can easily tip the scales in the trade-off between research needs and expected benefits of using data-enabled objects: As research-specific artifacts, these objects often do not have (much) end-user utility, and they present a potential case of ethical ambiguity~\cite{Zuboff_2015_JIT}. In short, the benefits to design research need to be weighed against the risks of privacy violation. As a result, design researchers face a dilemma in employing data-enabled objects: Are the risks worth the benefits? 

To better design data-enabled objects for research, we need to explore how privacy design can address the tension between participant privacy and ethnographic needs. Apart from privacy guidelines and legislation pertaining to data protection, ideas about privacy remain highly subjective, varying from person to person~\cite{mulligan_2012}. Researchers have emphasized the need for designing a means of data control for people to explicitly address their specific privacy needs during data collection~\cite{Cheng_2019_DIS, Pierce_2019_rtd}. However, data control cannot fully address all privacy issues, as people may not always be aware of all potential privacy issues, and, furthermore, some privacy issues arise unexpectedly and thus can be difficult to address beforehand~\cite{Pierce_2019_CHI}. 
A comprehensive investigation into everyday privacy issues related to data collection is necessary.

This investigation can begin to explore ``what method is appropriate for a given situation''~\cite{mulligan_2012} by examining the observations of individuals in that situation. By creating in-situ experiences of living with data-capturing devices and studying the real reactions that arise, participants and design researchers can better examine and reflect on their own privacy requirements during the experience. For example, a tool such as \textcolor{minor}{S}ensorstation~\cite{denefleh_2019_DIS} \textcolor{minor}{provided participants with a means to design and deploy sensors in their shared apartments.} 
Such deployment allowed design researchers to investigate the actual changes in people's behaviour and gain new understandings \textcolor{minor}{about} design\textcolor{minor}{ing} for connected sensors. Therefore, this work uses a similar approach \textcolor{minor}{for} facilitating design researchers to design, build and deploy the data-enabled objects into the field. As such, we can capture diverse examples and reflections from their design practices. However, this work was not meant for examining privacy issues.  Instead, this work aims to provide a toolkit that can be used for supporting design researchers \textcolor{minor}{in} prototyp\textcolor{minor}{ing} privacy-aware data-enabled objects and explor\textcolor{minor}{ing} more different strategies \textcolor{minor}{for} building appropriate interaction\textcolor{minor}{s} with different people.

Our main method in this study is Research through Design~\cite{Koskinen_2011_book}, an iterative research exploration process in which the research inquiry is posed via a design implementation, and then re-framed and re-focused based on reflections on the results of the inquiry~\cite{ylirisku_2009_CHI}. However, in this process, the actual outcomes cannot be fully predicted or planned~\cite{Koskinen_2011_book}. Thus, researchers need to continually re-frame and re-focus their research questions through a reflection-in-action process~\cite{schon_1983}. To better present an overview of our research journey, our following research questions \textcolor{minor}{are} as follows:
\begin{enumerate}
\item What are potential privacy challenges in using data-enabled objects to conduct ethnographic research in the wild?
\item How can we support design researchers in applying a tool for designing and using context-specific data-enabled objects?
\item How will design researchers use the tool to explore and build different and common design strategies for data- enabled objects that will take into account needs for  privacy?
\item How might our understanding of designing for privacy change through the process of privacy-aware design exploration and in-the-wild data-enabled research practice?
\end{enumerate}

We aim for the following contributions to the Design and HCI communities. First, we \textcolor{minor}{will} identify design challenges in designing and working with data-enabled objects. Second, we \textcolor{minor}{will} present an open-source toolkit, the Connected Peekaboo Toolkit, for design researchers to use in exploring, prototyping and testing their data-enabled objects using various sensor modalities and tailoring them to specific contexts. As such, this toolkit enables a controllable intervention that allows design researchers to conduct a bottom-up design space exploration for data-enabled objects in the wild. Third, we \textcolor{minor}{will} give a detailed account of how the toolkit was used by 18 design teams of 4-5 undergraduate students and, in so doing, identify opportunities and future directions for the development and use of data-enabled objects. Finally, the results from the teams have allowed us to identify that privacy is not an obstacle, but rather can be a driver in facilitating a shared agency between researchers, participants, and data-enabled objects.

\section{Background: Research through Data-Enabled Objects}
\begin{figure}
 \centering
 \includegraphics[width=1\linewidth]{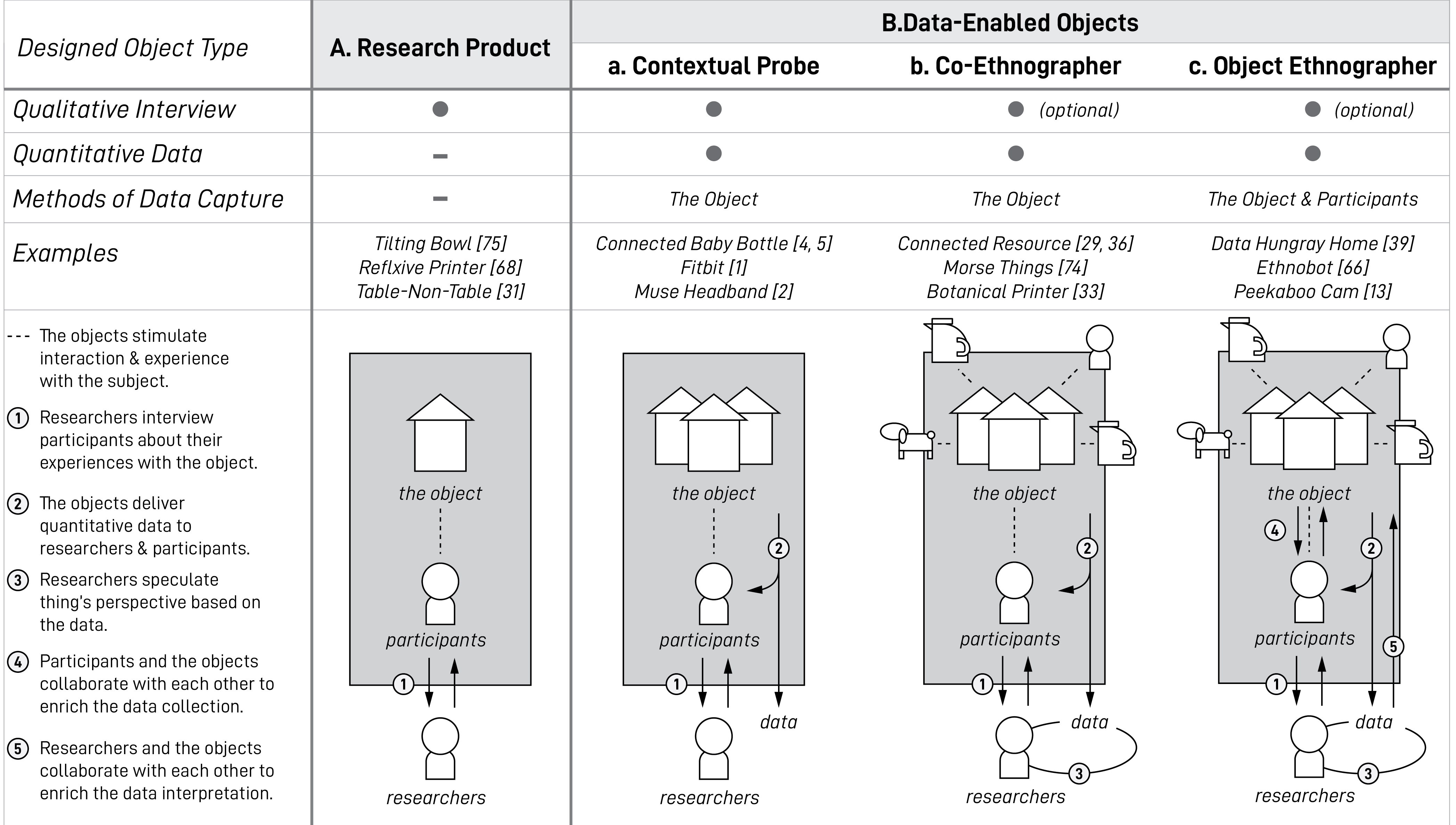}
 \caption{Prior design research \textcolor{minor}{has} deployed various purpose-built research artifacts into the field and studied the interactions or non-interactions with the artifact. This figure shows a comparison of different uses of artifact\textcolor{minor}{s} in prior work\textcolor{minor}{, as} (A) Research Product~\cite{Odom_2016_CHI} and (B) Data-Enabled Object. (A) Research Products are deployed \textcolor{minor}{to} stimulate novel experiences and interactions with participants; thus\textcolor{minor}{,} researchers can interview the participants about \textcolor{minor}{these} novel experiences (see \textcolor{minor}{\textcircled{1}}). (B) Data-enabled objects not only extend the qualities of the objects but also deliver quantitative data for researchers to \textcolor{minor}{use to} understand the novel experiences both from the participants' interview\textcolor{minor}{s} and \textcolor{minor}{the} data \textcolor{minor}{(see \textcircled{2})}. \textcolor{minor}{The data can be used for not only understanding participants but also for inspiring researchers to speculate on the thing's perspectives (see \textcircled{3}). At the same time, the more agency given to the data-enabled objects to collaborate with participants and researchers for capturing data (see \textcircled{4} and \textcircled{5}), the more it could change the way that people perceive data-enabled objects, recognizing them as taking on different roles, from probe to agent (a. contextual probes, b.co-ethnographers, c. object ethnographers).}}
  ~\label{fig:researchobject}
\end{figure}

In this section, we portray research objects that have been built by design researchers and how these objects shift their agency when being enriched with more data-enabled capabilities. \textcolor{revision}{We selected the examples based on whether the research objects were purposely-built, were of sufficient quality, and played major roles in capturing and delivering data for further analysis. These research objects can be used for provocatively intervening and capturing data on the new interactions or passively capturing data on existing interactions. However, they are similar in that they all deliver data for design researchers to use as part of the research-through-design (RtD) process. Based on Thing Ethnography~\cite{Giaccardi_2016_DIS}, we analysed the examples by identifying how different actors (research objects, participants and researchers) collaborated and interacted with each other during the design ethnography process (see the interactions between research object, participants, and researchers identified in Figure~\ref{fig:researchobject}).}

\subsection{\textcolor{minor}{From Research Products to Data-Enabled Objects}}
There is a long trajectory in design research that deploys self-designed research artifacts into the field and studies the interactions that emerge with the artifact, laying a groundwork for future design directions. We illustrated different methods of deploying research artifacts in Figure~\ref{fig:researchobject}. For example, in Figure~\ref{fig:researchobject}-A, the researchers built \textit{research products};
 these products possess well-designed and finished qualities that are meant to prompt participants into a natural engagement with the object~\cite{Odom_2016_CHI}. In \textcolor{minor}{F}igure~\ref{fig:researchobject}-A, we note several examples of such products: Table-non-table ~\cite{hauser_CHI_2018}, Tilting bowl~\cite{Wakkary_2018_CHI}, 
 and Reflexive Printer~\cite{tsai_2014_DIS}. All of these are objects with finished qualities that create a state of ``material speculation'' ~\cite{wakkary_2015_CA}, i.e., they make it easy for people to believe that they are seeing themselves as users interacting with the objects. This allows design researchers to study the novel experiences stimulated by the products~\cite{Gaver_2004_Interactions} and to investigate people's reactions and reflections on using the products through interviews. In employing the research products in studies, these design researchers combine\textcolor{minor}{d} design practice and ethnographic approach\textcolor{minor}{es} for \textit{design ethnography}, which is ``a set of data collection and \textcolor{minor}{analysis} perspectives, assumptions and skills that can be used effectively and efficiently to understand a particular environment or domain of people for the express purposes of designing new technology products''~\cite{salvador_1997_CHIEA}. 

To extend the design ethnography, researchers have increasingly built their research products to be equipped with sensing and connected technology (see \textcolor{minor}{F}igure~\ref{fig:researchobject}-B). For example, Bogers et al. designed the Connected Baby Bottle which is \textcolor{minor}{an} everyday baby bottle that can capture the data when users feed their babies~\cite{Bogers_2016_DIS, Bogers_2018_DIS}. Wakkary et al. built a set of connected bowls, called Morse Things, which capture and communicate the data in-between the bowls when interacting with their users~\cite{wakkary_2017_DIS}. These objects essentially have finished product quality
and, at the same time, capture and deliver in-situ data to the researchers for further investigation. We call these objects~\textit{Data-Enabled Objects}\textcolor{minor}{,} referring to object\textcolor{minor}{s} which can capture and deliver data for further uses. The data-enabled objects can be used as a research product that provides data perspectives and support design researchers in studying activities and behaviors not only from people's self-reported experiences but also from statistical and other quantitative contextual data. 

To better define \textcolor{minor}{how} data-enabled objects can be used in design research, we further examine and explore \textcolor{minor}{in Figure~\ref{fig:researchobject},} how the emerging data-enabled capabilities change the perspective of researchers, from using the research objects \textcolor{minor}{as} a contextual probe to \textcolor{minor}{using it as an active agent capturing data.} This examination allows us to identify the design challenges for privacy-aware data-enabled objects. 

\subsection{Using Data-enabled Objects as Contextual Probe\textcolor{minor}{s}}
Design researchers can use data-enabled objects as contextual probes that participants not only interact with but, additionally, capture data through these interactions and from the environment. For example, prior design researchers have built and used various sensor-augmented research objects such as Connected Baby Bottle~\cite{Bogers_2016_DIS, Bogers_2018_DIS}, The Living Room of the Future~\cite{sailaja_2019_TVX}, or various self-tracking devices (e.g., Fitbit~\cite{fitbit}, Muse headband~\cite{muse}) to capture in-situ data from the context of deployment. \textcolor{revision}{Connected Baby Bottle~\cite{Bogers_2016_DIS, Bogers_2018_DIS}, a baby bottle embedded with sensors, was used to capture in-situ data about how parents fed their babies. The Living Room of the Future~\cite{sailaja_2019_TVX} was a showroom that incorporated with both real and fictional technologies that collected real-time data from the audience and invited them to experience how data potentially will flow in the future living room. Several researchers also used wearable tracking devices to collect physiological data that could be used for improving sleep quality~\cite{nguyen_2020_nordichi} or treating alcoholism~\cite{you_2019_mobileHCI}.} \textcolor{minor}{In these cases, the} additional data-capturing capability \textcolor{minor}{provided researchers with data for studying the participants from a more objective perspective (e.g., quantified self~\cite{luption_2016_book}).} Armed with these in-situ data, design researchers can also investigate multiple contexts remotely~\cite{bolt_2010} and conduct interviews with participants \textcolor{minor}{in order} to interpret the data together, and to understand the actual contexts.
We identify that these lines of research objects suggest a potential strategy of using data-enabled objects as \textit{contextual probes} that use data as a stimulus to probe participants for further interpretation (see Fig~\ref{fig:researchobject}Ba-1) and provide an alternative data perspective about the participants (see Fig~\ref{fig:researchobject}Ba-2).

\subsection{Using Data-enabled Objects as Co-Ethnographers}
Design researchers can use data-enabled objects as co-ethnographers to go further than just investigating the in-situ contexts. With this approach they can also speculate on alternative perspectives inspired by the data. For example, Giaccardi, et al.~\cite{Giaccardi_2016_DIS} described their research objects as ``co-ethnographers'' because their objects contributed the first-thing view of the environment, enriching design ethnography with broader perspectives that can help discover more-than human-centered experiences. \textcolor{revision}{Data-enabled objects as co-ethnographers represents a special case of ``connected things~\cite{giaccardi_2019}'' working in partnership with humans during design ethnographic research.}
Other design researchers also used their research objects such as Morse Things~\cite{wakkary_2017_DIS} and Botanical Printer~\cite{hsu_2018_DIS}, to not only stimulate novel experiences with their participants but also to interpret and speculate on the thing's perspective. \textcolor{revision}{For example, Morse Things~\cite{wakkary_2017_DIS} was designed to create a data flow between connected bowls and cups using Morse code; the interactions can be followed online, and are thus presented as an invitation to people to speculate on the objects' hidden communications. Botanical Printer~\cite{hsu_2018_DIS} translates environmental data (e.g., CO\textsubscript{2}) into different narratives that engage people in a slow, interactive experience somewhat like the experience they might have with plants.} In these cases, researchers can interpret more possible scenarios beyond their human perspectives, where the data is no longer only about the interaction with the immediate end users~\cite{Giaccardi_2020_chapter}. They can also capture interactions entangled with other users and stakeholders~\cite{MurrayRust_2019_HTTF}, as well as imperceptible communications between objects~\cite{Nicenboim_2018_CHI_EA, lindley_2020_DIS}, and even interactions between non-human players, such as pets and a cleaning robot~\cite{Coulton_2019_desigjournal}. \textcolor{revision}{Another example is Connected Resource~\cite{Nicenboim_2018_CHI_EA, Giaccard_2018}, which is a set of connected products designed for the elderly and inspired by the thing-centred ethnographic data that shows how older people adapt existing objects to support their daily routines. The data collected through these connected products have supported designers in exploring interactions hidden from people and, by doing so, have helped them identify still more new design opportunities and design connected products (e.g., Connected Stone, Connected Bell, and Connected Magnets).} 


From their work, we identify that data-enabled objects can be used as co-ethnographers, \textcolor{minor}{capturing} data about human\textcolor{minor}{s} and other social things for facilitating researchers to be engaged \textcolor{minor}{with} a data-enabled interpretation space (see Fig\textcolor{minor}{ure}~\ref{fig:researchobject}Bb-3). Such a data-enabled interpretation space sheds new light on design ethnography that can lead toward the vision of \textit{More-Than Human-Centred Design Thinking}\textcolor{revision}{~\cite{coulton_2019_designjournal, frauenberger_2019_CHI, wakkary_2021_book, Giaccardi_2020_chapter}}.

\subsection{Using Data-enabled Objects as Object Ethnographers}
Design researchers can use data-enabled objects as object ethnographers to not only capture and deliver data perspectives (the same as when using objects as contextual probes or co-ethnographers) but also to autonomously support the data-collection process. Here, we do not mean that the data-enabled objects would replace the job of human ethnographers. Instead, we are claiming that these data-enabled objects can be more than just passive data collectors; they can take on an active role participating and influencing the data collection process. Data-enabled objects can also be designed with autonomous capabilities working as a stand-alone device, interacting with the research subjects during the ethnographic study. For example, Tallyn et al. developed an Ethnobot that can actively probe and collaborate with the participants to capture data in the field~\cite{tallyn_2018_CHI}. Lee-Smith et al. proposed Data Hungry Home where various objects can have different intentions \textcolor{minor}{for using} and \textcolor{minor}{conducting} and \textcolor{minor}{intervening in} the data collection~\cite{smith_2019_HTTF}. Cila et al. proposed three different object roles (e.g., collector, \textcolor{minor}{actor}, or creator) which can use data to create different experiences in the field~\cite{Cila_2017_CHI}. Cheng et al. designed Peekaboo Cam\textcolor{minor}{,} which emphasise\textcolor{minor}{s} designing an interaction between the research objects and the participants to negotiate the appropriate data flow sharing~\cite{Cheng_2019_DIS}. 

In these examples, we identify that data-enabled objects can become \textit{object ethnographers} that are not only delivering data but also actively intervening \textcolor{minor}{in} data collection with participants and researchers (see Fig~\ref{fig:researchobject}Bc-4 \& 5). Data-enabled objects as object ethnographers build diverse relationships with various stakeholders during the course of a study (e.g.,with pets, houseplants, or acquaintances of the participants).

No matter how these sensor-augmented objects are built, whether as a passive contextual probe or as an active agent, we see that they are all purposefully designed and uniquely identifiable as a stand-alone device. The objects are integrated with localized sensing that \textcolor{minor}{can} perceive events and human activities occurring in the physical surroundings close to the objects. Additionally, data-enabled objects \textcolor{minor}{have} open accessibility that provide\textcolor{minor}{s} a data interface \textcolor{minor}{with which} design researchers or for other objects \textcolor{minor}{can} access and use the collected data. These objects interact with people and the surroundings, and their interactions \textcolor{minor}{can} influence the data collection and output. The big difference between each type is that the data-enabled objects can be programmed \textcolor{minor}{to} actively or passively influence the ethnographic results. However, such capability also increases the complexity of designing the interactions and relationships among the participants, the data-enabled objects, and the researchers. Although in most prior studies the researchers were able to successfully deploy their data-enabled objects into the wild, their work over-simplified the provisions for ethical data-gathering, for example, by only providing an information letter and consent form for participants to read and sign. In view of emerging legal data-protection frameworks, \textcolor{minor}{this may have been} enough for the past, but it seems hardly enough to fully protect participants' rights and their privacy in nowadays. Considering that a new ethnographic field can be shaped by using data-enabled objects, researchers need to emphasize investigating and proposing new approaches for conducting their studies~\cite{crabtree_2006}. This calls for an exploration of the potential challenges of employing data-enabled objects in the field.

\section{Privacy Design Challenges with Data-Enabled Objects}

To explore how data-enabled objects can better perform their tasks in capturing contextual data and interacting with people in the wild, we will next present speculation and discussion on the possible design challenges involved. \textcolor{revision}{Based on research on human ethnography~\cite{hammersley_2007_book, Jones_Smith_2017}, ethics of data collection~\cite{mortier_2014, limerick_2014, Shneiderman_2004_book, Nippert_2007_ijd, Pierce_2019_rtd}, and ethnographic practice with things (e.g. data-enabled objects)~\cite{Cheng_2019_DIS, Giaccardi_2016_DIS, tolmie_2010, lovei_2020_DIS}, we identified three challenges—Engagement, Empowerment and Enactment. These three challenges outline how data-enabled objects can play a role in an ethnographic study, including in the processes of deployment, data collection, and data interpretation, all in collaboration with participants and researchers.} 
If we take the example of the Peekaboo Camera~\cite{Cheng_2019_DIS}, we can see that, in each phase, researchers need to plan where and how to deploy their sensor-augmented research prototypes into the field where \textcolor{minor}{they} can capture the relevant data that matches their research interests. At the same time, researchers need to consider ways of preventing the deployment to interfere with participants' privacy. Furthermore, as different situations can arise in the field, an appropriate manner for their research prototypes \textcolor{minor}{to conduct} a long-term ethnographic practice need\textcolor{minor}{s} to be explored. 

Therefore, we discuss the potential design challenges during the design ethnographic practice\textcolor{minor}{,} starting with three major questions--- (1) How do data-enabled objects \textbf{engage} \textcolor{minor}{in} everyday activities to not only capture relevant contextual data but also to adapt to a changing context? (2) How can data-enabled objects \textbf{empower} different stakeholders with a sense of control that will match their expectations and requirements? (3) How do data-enabled objects better \textbf{enact} their autonomous capabilities for engagement and empowerment as contexts change? 

This analysis allowed us to better identify what type of toolkit we needed to build to support designers in using and designing data-enabled objects that can allow for designing for individual privacy needs. 
 
\subsection{Engagement: Capturing Contextual Data Passively or Actively}
\textit{How do data-enabled objects engage into everyday activities to not only capture relevant contextual data but also to adapt to a changing context?}

Data-enabled objects capture the contextual data of the local context in which they are located. Therefore, researchers need to identify how the location is relevant to the context they aim to study, and how they can integrate data-enabled objects into the everyday practice of the participants or how the objects can engage in people's activities in order to capture relevant data. However, this seamless integration into everyday practice can carry the risk of the interactions and growing relationships with the participants turning the data-enabled objects into an ominous black box that causes the participants to experience anxiety. On the one hand, while researchers have emphasised the need of shaping the human-data interaction through giving participants legibility, agency and negotiability to intervene in the data collection, too many interventions may influence the participants' daily activities. As a result, the data-enabled objects \textcolor{minor}{that allows} too many interventions cannot capture \textcolor{minor}{data that reflects} the natural behaviours \textcolor{minor}{of} the participants. Therefore, data-enabled objects should have \textcolor{minor}{a} quality of engagement \textcolor{minor}{that,} like \textcolor{minor}{a} human ethnographer\textcolor{minor}{,} know\textcolor{minor}{s} how to ``live and work'' with the participants~\cite{hammersley_2007_book}.
For human ethnographers, to ``live and work'' in the environment they are studying (i.e., observing and collecting data in the wild) is an important means for observing participants' natural engagement in the wild~\cite{hammersley_2007_book}. 
This requires the ethnographer to refrain from being a shadowy or virtual presence that secretly observes people from behind the stage. Instead, they must engage with the participants and occupy space in the context and everyday life of the participants. However, the ethnographer must also take care not to be an interventionist by disrupting existing routines and thus influencing their observational results as well as potentially invading someone's privacy. They also must be aware of the social norms of the participants, as this will help them to know when to ``step back and re-state the boundaries of the relationship''~\cite{Jones_Smith_2017}. By adopting the ethnographer's spirit of ``live and work'', we can see that data-enabled objects also need to acquire the capability of adapting their engagement in the process of capturing data as the situational context shifts and changes. 
A design challenge for designers is to translate \textcolor{minor}{this} quality of \textcolor{minor}{living} and \textcolor{minor}{working with participants} into data-enabled objects, thus allow\textcolor{minor}{ing} the objects to have \textcolor{minor}{a} contextual awareness \textcolor{minor}{that will give them the ability to perceive} when to step in and step out of the data collection process without disrupting their existing routines. 


\subsection{Empowerment: Enabling a Sense of Control for People}
\textit{How can data-enabled objects empower different stakeholders with a sense of data control that will match their expectations and requirements?}

Data-enabled objects, as computational devices with on-board memory and processing capabilities, can capture long-term data consistently without fatigue; yet, the objects can be insensitive to the context in which they are operating and to changes therein. As a result, researchers have examined the importance of empowering people with the ability to intervene in a data-collection system, by incorporating legibility, agency, and negotiability into the data-collection process~\cite{mortier_2014}. 

From examining their work, we derive the requirement that the data-enabled objects themselves should give participants support \textcolor{minor}{in understanding} the purpose of the data collection. Additionally, the objects should provide participants the agency to intervene in the data collection, and allow them the negotiability of always being able to evaluate their decisions as contexts change. In short, data-enabled objects should empower participants to have a sense of control during the course of a study. This sense of control should be understood as more than \textcolor{minor}{a} button design that allows people to directly trigger, modify, or intervene in the data capture. Instead, \textcolor{minor}{it should require} designers to take an individual's state of mind into design consideration. For example, the participants should feel that ``they are in charge of the system and that the system responds to their actions''~\cite{limerick_2014, Shneiderman_2004_book}. Therefore, the basis of data control cannot be coded simply by using a well-written rule of distinguishing private or non-private data. Rather, it is a subjective and dynamic negotiation process of participants or users deciding~\cite{Seigneur_2004_trust, vanHeerde_2006_SSC} between what should be inaccessible and accessible, as their criteria for determining such decisions can change depending on different contexts~\cite{Nippert_2007_ijd}. For example, in design cases involving curtains~\cite{Pierce_2019_rtd} and controllable enclosures~\cite{Cheng_2019_DIS}, the designers built dynamic physical obstructions that allowed users to instantly control data being gathered by a camera. While building an interface or means of data-control interaction for participants can be a practical starting point, other design aspects which can influence participants' sense of control (i.e., form~\cite{Cheng_2019_DIS}) need to be explored.

\subsection{Enactment: Designing Automated Capabilities of Data Processing  }
\textit{How can data-enabled objects better enact their autonomous capabilities for engagement and empowerment as contexts change?}

Data collection is the key activity of data-enabled objects: They should be able to reliably capture relevant data from the field over a long period of time~\cite{tolmie_2010, lovei_2020_DIS}. To ensure reliable operation and high-quality data, 
researchers need to carefully design the automated data-processing capabilities of data-enabled objects. First, data-enabled objects require a relatively stable structure for storing and transferring data and other information during the course of the study. Second, while data-enabled objects can capture structured data under most conditions, this can potentially lead to noisy data, as the data-enabled objects cannot selectively tune into what will ultimately be considered interesting for data analysis. Noisy data can be difficult for researchers as well as participants to interpret and make sense of. This, in turn, bears the risk that participants will not agree to the data collection as they cannot understand how the data might be further used and interpreted. For example, intimate details and behavior can be hidden within noisy data, and just the suspicion of this might lead a participant to abort their involvement in a study. To better conduct studies using data-enabled objects, researchers need to address the fundamental challenge of how to design an object's data-processing capabilities in a way that can maintain data quality and ensure desirable outcomes while protecting the privacy of participants throughout the course of the study.

Data-enabled objects can capture data consistently without fatigue, and they have a large enough memory to record data in almost arbitrary quantity and depth.
However, these strengths can also be weaknesses if data-enabled objects have inadequate flexibility in responding to the changing needs of participants and researchers as different situations arise in context and study focus. For example, data-enabled objects need to be programmed to know when to engage and disengage in people's activities, how to enable people to experience a sense of control over the data being gathered, and how to process the raw data into meaningful data for further use. The three design challenges, \textit{engagement, empowerment}, and \textit{enactment} give us a structure to identify the potential limits and opportunities for developing a framework or a toolkit to support design researchers \textcolor{minor}{in} deploy\textcolor{minor}{ing} and us\textcolor{minor}{ing} data-enabled objects in \textcolor{minor}{a} research context. 

\section{Designing an Open Toolkit for Building Data-Enabled Objects}
Based on the three design challenges, engagement, empowerment and enactment, we have developed an open toolkit that supports design researchers \textcolor{minor}{in} design\textcolor{minor}{ing} and \textcolor{minor}{implementing robust} data-enabled objects \textcolor{minor}{that address} these challenges. As such, we can examine their design practices and investigate \textcolor{minor}{the} outcome\textcolor{minor}{s} for unfolding the privacy design space for data-enabled objects.

\subsection{Toolkit specification}
The purpose of the toolkit is to support design researchers \textcolor{minor}{in} build\textcolor{minor}{ing} and explor\textcolor{minor}{ing} possible privacy design spaces for data-enabled objects. In revisiting the three design challenges, the corresponding requirements for the toolkit can be listed as follows: 
\textit{Toolkit design requirements for Engagement} 
\begin{itemize}
\item The toolkit should provide a fundamental data infrastructure that allows design researchers to build, design, deploy, and test their data-enabled objects with participants in the wild.
\item The toolkit should provide a flexible design template that allows design researchers to design the data capturing interaction with their participants. 
\end{itemize}

\textit{Toolkit design requirements for Empowerment}
\begin{itemize}
\item The toolkit should provide an interaction design that gives participants a sense of control during the ethnographic practices.
\item The toolkit should gather quality data with a high degree of legibility for their participants and for design researchers. 
\end{itemize}

\textit{Toolkit design requirements for Enactment}
\begin{itemize}
\item The toolkit should provide flexibility \textcolor{minor}{in supporting design researchers in negotiating with their participants regarding the deployment plan (e.g., location).}
\item The toolkit should allow data-enabled objects to have contextual sensing for detecting changing contexts. 
\end{itemize}

\subsection{Connected Peekaboo Toolkit}
\begin{figure*}
  \centering
  \includegraphics[width=\linewidth]{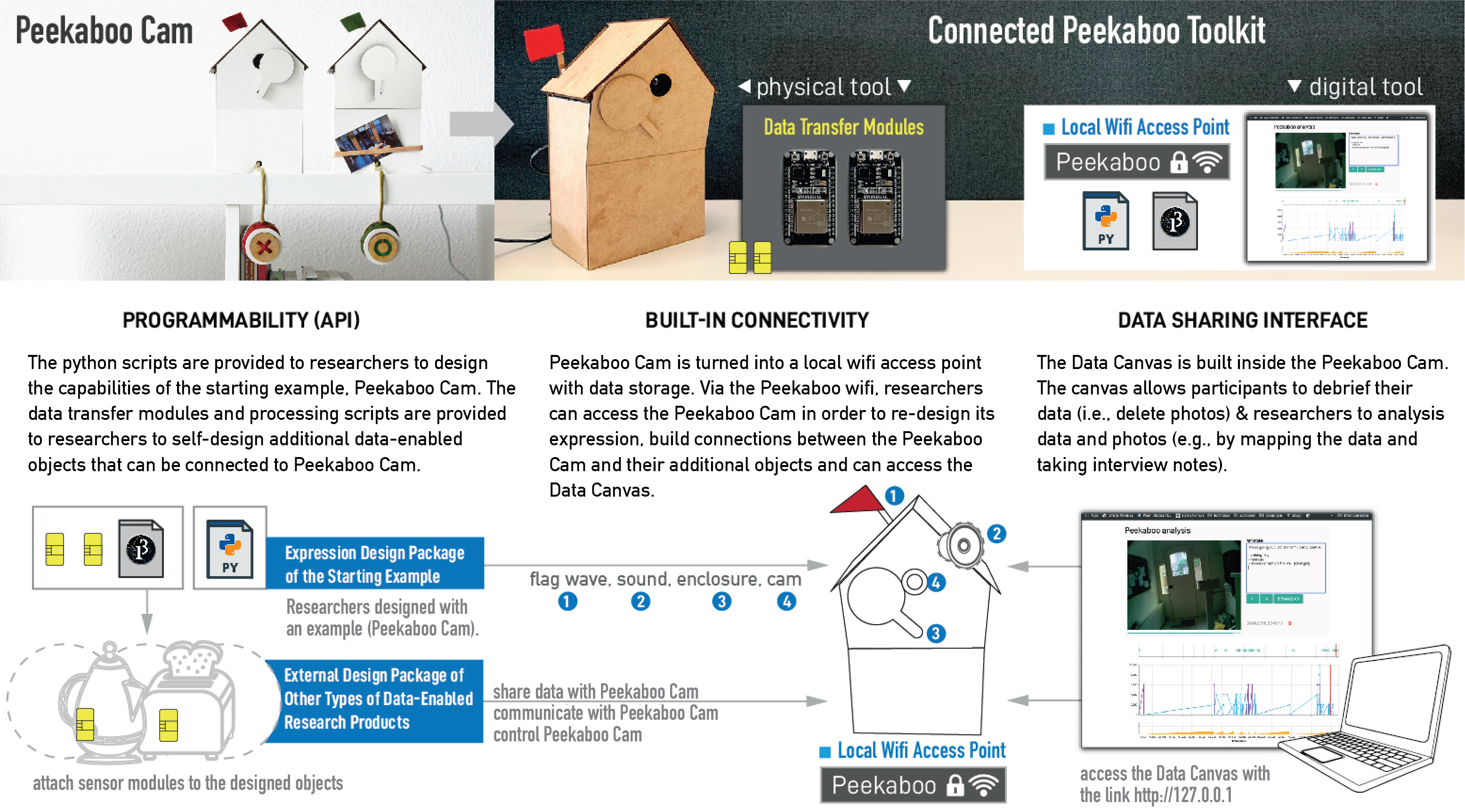}
  \caption{The Connected Peekaboo Toolkit (CPT) is built on an existing example, Peekaboo Camera~\cite{Cheng_2019_DIS} which is an automatic time-lapse camera with an interactive data control button. \textcolor{minor}{The} CPT turns the Peekaboo Camera into an open design\textcolor{minor}{, so that} designers \textcolor{minor}{can} flexibly design its expression. We call the camera \textit{the CPT camera} in this paper. The CPT camera contains a wifi router\textcolor{minor}{,} which allows design researchers to flexibly connect with multiple sensing objects for wider data collection locally. The CPT camera also \textcolor{minor}{provides} data storage and an additional data sharing interface (Data Canvas)\textcolor{minor}{,} which allows design researchers and participants to efficiently \textcolor{minor}{conduct} de-brief\textcolor{minor}{ings all of} the captured data.}~\label{fig:toolkit}
\end{figure*}

Based on \textcolor{minor}{these} specification\textcolor{minor}{s}, we present the Connected Peekaboo Toolkit (CPT), \textcolor{minor}{which} design researchers can use to design a connected system of data-enabled objects based on \textcolor{minor}{the} solid foundation of a working prototype with built-in connectivity, programmability (APIs), and a data-sharing interface. We present these three features in Figure~\ref{fig:toolkit}: Design researchers can, with little effort, add additional sensors and actuators, thereby changing the programmed behavior and data-collection mechanisms. Overall, the design of \textcolor{minor}{the} CPT went through several iterations, starting from re-building the original design called Peekaboo Camera~\cite{Cheng_2019_DIS} (see the original camera in Figure~\ref{fig:toolkit}). This prior study describes a camera-based, interactive object that was deployed and tested with two families. In a nutshell, the study resulted in rich contextual data and explorations of privacy design strategies. However, the researchers still encountered difficulties in prototyping and implementing data-enabled objects, such as how to find a balance between privacy and ethnographic needs, how to establish trust with participants, and how to exploit the various privacy design opportunities in even just a single home. Here we see an opportunity \textcolor{minor}{in inviting} student design researchers to re-design the existing privacy-aware data-enabled object and explore more design spaces beyond what the prior work has explored. 

\subsubsection{From Peekaboo Camera to an Open Design Toolkit}
We adapted the essential four research processes outlined in the Peekaboo Camera work into the toolkit design: deployment, data capturing, data debriefing, and interviewing \& interpretation~\cite{Cheng_2019_DIS}. The goal was to design the features that would support the student teams in conducting each phase \textcolor{minor}{of} their studies. 
Second, we simplified the mechanics and replaced the earlier interaction means by providing functional stubs that could be implemented by the student teams. The software was made available as open-source, with several iterations that fine-tuned the data-logging capabilities and the startup behavior. Another iteration was necessary to extend and refine the code documentation and how-to guides for the student teams. A final iteration focused on the browser-based interaction with the collected data and the annotation thereof. Overall, we went through a fast-paced design process in the weeks leading up to the course, and even after the course started, a few modifications to the code documentation were made because some teams inquired about modifying deeper technology layers than originally provided for. The following presents the important features of the CPT.

\subsubsection{Built-in Connectivity for Flexible Deployment \& Wider Data Collection}
\begin{figure*}
  \centering
  \includegraphics[width=\linewidth]{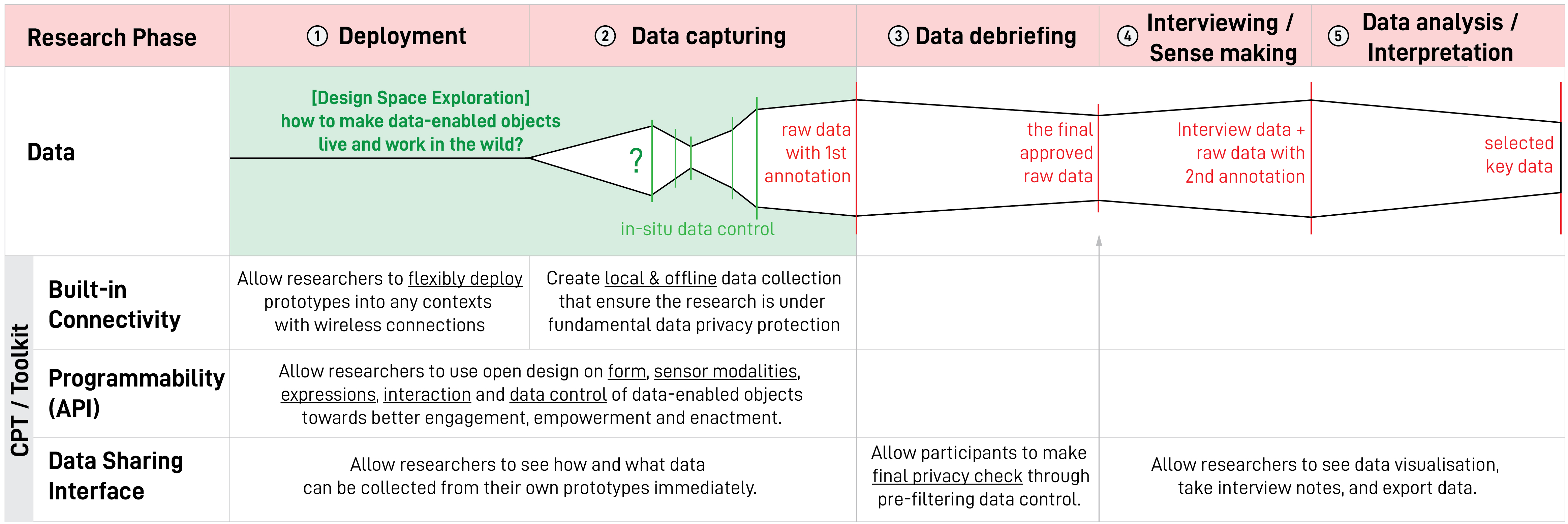}
  \caption{Overview of how three design features (built-in connectivity, programmability (API), and data sharing interface) support researchers \textcolor{minor}{in} explor\textcolor{minor}{ing} the interaction design for data-enabled objects.}~\label{fig:toolkitprocess}
\end{figure*}

The toolkit provides \textit{built-in connectivity} in the form of a separate WiFi network, which has distributed sensors and actuators that can be used to send and receive data. This ensures that data from the local ecosystem of connected devices will not be shared externally. This feature support\textcolor{minor}{s} the research phase\textcolor{minor}{s} of deployment and data capturing illustrated in Figure~\ref{fig:toolkitprocess}. In these two beginning research phases, the \textit{built-in connectivity} \textcolor{minor}{not only gives} researchers more flexibility to deploy their prototypes in any location but also make\textcolor{minor}{s} sure all (research) data is stored on the CPT camera. The latter point is important because it can establish trust in the participants that data collection and processing practices are being undertaken responsibly\textcolor{revision}{.} When it comes to research ethics and privacy legislation for data control, local connectivity is the required infrastructure for securing all data in a local, offline space (in addition to other measures). As a side effect, local connectivity allows the data-enabled objects to be a stand-alone setup that can be deployed quickly and flexibly into any home environment. The design researcher will not need access to WiFi networks or a home-computing infrastructure. Technically, we disabled Internet access in the CPT camera, re-routed the DNS resolution and HTTP requests to the internal hub \textcolor{minor}{server} and ensured that all aspects of the CPT camera would operate as an offline stand-alone artefact. Through using the open-source connectivity platform OOCSI~\cite{funk_mathias_2019_1321220} and configuring a local WiFi access point with its own SSID and password, the CPT camera contains the infrastructure to easily connect sensors and actuators. We also provided example code on the wireless sensor modules (based on the ESP32 platform), which design researchers can use to implement and connect sensing objects allowing wireless data transfer to the CPT camera. 

\subsubsection{Programmability (APIs) for Open Interaction \& Expression Design}
To explore appropriate interaction styles for data-enabled objects, the CPT provides programmability (APIs), which enables access to its main functions, signaling and taking photos. Design researchers can re-design the behavior and expression of their data-enabled objects to explore more appropriate interaction styles in their chosen contexts, e.g., behaviors and interactions can be made more responsive or more obtrusive during the deployment and data capturing phase\textcolor{minor}{s} illustrated in Figure~\ref{fig:toolkitprocess}. In practice, design researchers can use the existing code on the CPT camera to change how their data-enabled objects collect data, and can also control how notifications are expressed on the CPT camera. This functionality is also replicated in an open API. We provide a control software based on Python (https://www.python.org/) and Processing (https://processing.org/) that shows how to trigger different data-enabled object functions via the API during the times when the object is taking photos. For example, a designer could add an external module that senses human presence and then modifies the object's internal behavior to deactivate the photo-taking function for a few minutes. Both the internal code and the API are well-documented, so that design researchers can effectively expand the interaction of data-enabled objects with the environment.

\subsubsection{Data-Sharing Interfaces for Data Debriefing \& Co-Interpretations}
To ensure that only data that are approved by participants will be visible for design researchers (i.e., student designers, in our case), we designed a data-sharing interface, a web-based \textit{data canvas}, that visualizes all collected data and every captured photo on a timeline. The canvas allows participants to browse the data by themselves and delete data that they do not want to share (regardless of the reason). Participants can easily connect to the CPT camera via the built-in WiFi network and browse the data canvas with a web browser on their own computers. The data canvas is essentially a tool for participants to self-filter and perform data control. After the pre-filtering, the data canvas assists both participants and student designers in data sense-making for further co-interpretation. The data sharing interface can support the researchers and research participants \textcolor{minor}{in} examin\textcolor{minor}{ing} all of the current data during \textcolor{minor}{all of the} research phase\textcolor{minor}{s} including deployment, data capturing, data debrief\textcolor{minor}{ing}, interview\textcolor{minor}{ing} and data analysis\textcolor{minor}{, as} illustrated in Figure~\ref{fig:toolkitprocess}. It is possible to get an overview of all the data, to spot patterns in the data, and to annotate findings or activities directly on all collected data and photos.

\subsection{Summary}
The Connected Peekaboo Toolkit, as a camera-based, connected, open, highly extensible and scalable design platform, forms a starting point for designing and working with data-enabled objects. Given the features of built-in connectivity, programmability, and data-sharing interfaces, designers and researchers can re-design and tailor the behavior and environment awareness of their data-enabled objects for different contextual studies. More details of \textcolor{minor}{the} CPT can be found in the open-source repository at [https://github.com/FutureEveryday/ConnectedPeekabooToolkit]. Finally, our next step \textcolor{minor}{was} to provide \textcolor{minor}{the} CPT to multiple student design teams \textcolor{minor}{to use} in building their own data-enabled objects and conducting practice-based research.

\section{Design Explorations with 18 Design Research Teams}
\begin{figure*}
  \centering
  \includegraphics[width=\linewidth]{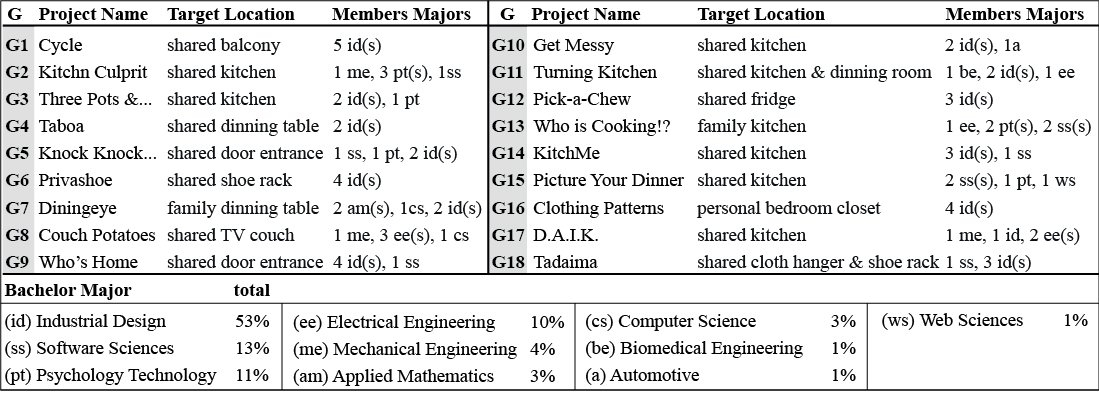}
  \caption{Our toolkit successfully allowed 18 design teams to customize their own data-enabled objects. This figure presents the 18 different types of data-enabled objects that were created and their target locations. The figure also presents the backgrounds of the 75 students.}~\label{fig:backgrounds}
\end{figure*}
We provided the toolkit to an undergraduate design course with 75 students over eight weeks \textcolor{revision}{in the Netherlands in 2019} (see demographics in the Figure~\ref{fig:backgrounds}). The course explored values and practices in the relationship between people and technology for a future connected society. The students, being aware of the technical challenges from an earlier course, were able to work with the provided technology, i.e., the Connected Peekaboo Toolkit. To ensure that they could efficiently design, build, and finally deploy prototypes into an environment, the students were asked to form interdisciplinary design teams, i.e., team members were to have different backgrounds and majors, including design, engineering, and sociology. Accordingly, different team members focused on different activities, such as product design, coding, building electronics, and interviewing end-users. \textcolor{revision}{The student teams were asked to study homes such as student-shared houses or Dutch family homes in the Netherlands (see details of each design team's choices that have studied and deployed can be found in Appendix 1)}. \textcolor{revision}{At the same time, the formal template of the consent form was provided to each design team so that they could modify it to suit their own research needs. They had to inform their participants of the type of data to be collected, the location of deployment, and the number of study days. Each study was conducted only after the participants had signed the consent form.}
In re-designing the object's data-capturing capabilities, the design teams were allowed to re-design its shape and add any sensors and actuators—all through the use of CPT. 

\subsection{Design Exploration Procedure}
\begin{figure}
  \centering
  \includegraphics[width=\linewidth]{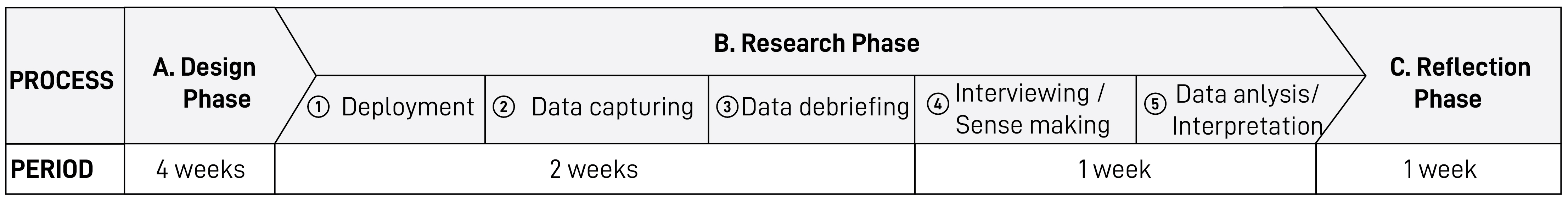}
  \caption{Design explorations with \textcolor{minor}{the} CPT \textcolor{minor}{consisted of} three main phases: A.\textcolor{minor}{D}esign, B.\textcolor{minor}{R}esearch and C.\textcolor{minor}{R}eflection. In detail, the research process include\textcolor{minor}{d} five stages: 1) \textcolor{minor}{D}eployment, 2) \textcolor{minor}{D}ata \textcolor{minor}{C}apturing, 3) \textcolor{minor}{D}ata \textcolor{minor}{D}ebrief, 4) \textcolor{minor}{I}nterview \& \textcolor{minor}{S}ense \textcolor{minor}{M}aking, and 5) \textcolor{minor}{D}ata \textcolor{minor}{A}nalysis \& \textcolor{minor}{I}nterpretation. After the explorations \textcolor{minor}{by the} 18 design teams, we received 18 prototypes and 18 design reports\textcolor{minor}{, which included} their design rationales, challenges, process, reflection\textcolor{minor}{s} and feedback on \textcolor{minor}{the} CPT.}~\label{fig:method}
\end{figure}

We illustrate our design process \textcolor{minor}{for} the course in Figure~\ref{fig:method}. As the figure shows, the process of design exploration \textcolor{minor}{using the} CPT consisted of three main phases \textcolor{minor}{covering a total of 8 weeks}: A. \textcolor{minor}{D}esign (4 weeks), B. \textcolor{minor}{D}esign \textcolor{minor}{E}thnography (3 weeks), and C. \textcolor{minor}{R}eflection (1 week). The design teams were asked to build their own data-enabled objects for a home observational study. To better guide the teams in designing and working with data-enabled objects, we introduced them to a general research process (B) with five steps: 1) \textcolor{minor}{D}eployment, 2) \textcolor{minor}{D}ata \textcolor{minor}{C}apturing, 3) \textcolor{minor}{D}ata debriefing, 4) \textcolor{minor}{I}nterviews \& \textcolor{minor}{S}ense \textcolor{minor}{M}aking, and 5) \textcolor{minor}{D}ata \textcolor{minor}{A}nalysis \& \textcolor{minor}{I}nterpretation. 



\subsection{Data Analysis}
After the entire process, we reviewed \textcolor{minor}{the} 18 different designs of data-enabled objects and 18 home studies undertaken using the data-enabled objects. The design teams each delivered a report documenting their entire process, including challenges and outcomes, as well as individual reflections and feedback on \textcolor{minor}{the} CPT  per team member (n = 75). 
To identify commonalities and differences in design rationales for designing privacy-aware data-enabled objects, 
all the design reports and reflection\textcolor{minor}{s} were coded using the inductive Thematic Analysis method, which provides us with a flexible research tool to collect, extract and group the data into meaningful subset\textcolor{minor}{s} based on our particular research interests~\cite{Braun_2006_QRP}. The detailed process is: The data were divided into analytic categories and the content was coded using \textcolor{minor}{a} set of themes and sub-themes created by three researchers. The coding and analytical procedure included three stages. First, one researcher (one of the authors) read through all the reports and annotated them with regard to: 1) design outcomes (what objects they had built), 2) rationales (why they wanted to build them), 3) reflections (challenges they faced and feedback they gained from their participants), and 4) feedback on \textcolor{minor}{the} CPT. Second, the researcher prepared and presented all of the annotated data to a group of experts with  mixed backgrounds in design and computer science (i.e., the rest of the authors). The researchers carefully examined the data, compared the initial themes with the raw data, and developed sub-themes. Last, the researchers (i.e., all the authors) combined and refined the sub-themes in an iterative, dialogic process until everyone agreed on them, and then the final themes were generated. The final themes represent a synthesis of how the participants (the student researchers) designed their data-enabled objects to be interactive and how they conducted their studies concerning privacy in the wild. The analysis allowed us to identify the design teams' commonalities and differences as a way to unfold the design space for data-enabled objects. 



\section{Findings from the Design Explorations}
\textcolor{revision}{With the CPT, the design teams deployed their prototypes in different types of homes (shared student houses, family houses, etc.) to capture visual data (photos in all groups) and non-visual sensory data (e.g. ultrasonic and infrared data for G1; pressure data for G18) during different periods of time. (See Appendix 1 for more details of the 18 design projects, including captured data types, total amount of visual and sensory data, deployed location, data collection period and design rationales)).} To give a general understanding of the types of data-enabled objects that the student design teams built, we have chosen two of the design projects to show in more detail. Second, we looked into all 18 design reports and reflections on building data-enabled objects using \textcolor{minor}{the} CPT, and identified five major design aspects \textcolor{minor}{of} the design space for privacy-aware data-enabled objects. These are: Form, Observational Perspective, Interaction \textcolor{minor}{Mode} of Data Capturing, Notification of Data Capturing, and Data Processing. Each one involves strategies for exploration and reflection on creating a balance between privacy concerns and ethnographic purposes for data-enabled objects concerning its engagement, empowerment and enactment in the wild. 

\textcolor{minor}{\subsection{Two Representative Examples from 18 Design Projects}}
In the study, we received 18 design projects exploring and prototyping data-enabled objects as enabled by CPT. We present an overview of these 18 projects in Figure~\ref{fig:backgrounds}, which shows that the 18 student design teams studied a variety of contexts in different households with multiple inhabitants, the households being either student-shared houses or traditional family houses. To look further into the details \textcolor{minor}{of} the system\textcolor{minor}{s} of data-enabled objects that \textcolor{minor}{the} student design teams built using \textcolor{minor}{the} CPT, the following section focus\textcolor{minor}{es} on two of the\textcolor{minor}{ir} design projects, Tadaima and Peek-a-boy. \textcolor{revision2}{We selected these two cases as examples because they represented two very different home contexts, a public context (i.e., a home entrance) and a private context (i.e., a bedroom). This allowed us to show how the design teams address privacy concerns in response to the different needs of participants in different situations.}

\subsubsection{Tadaima: Studying the Routines of People Entering a Home. }
\begin{figure}[h]
\centering
  \includegraphics[width=1\columnwidth]{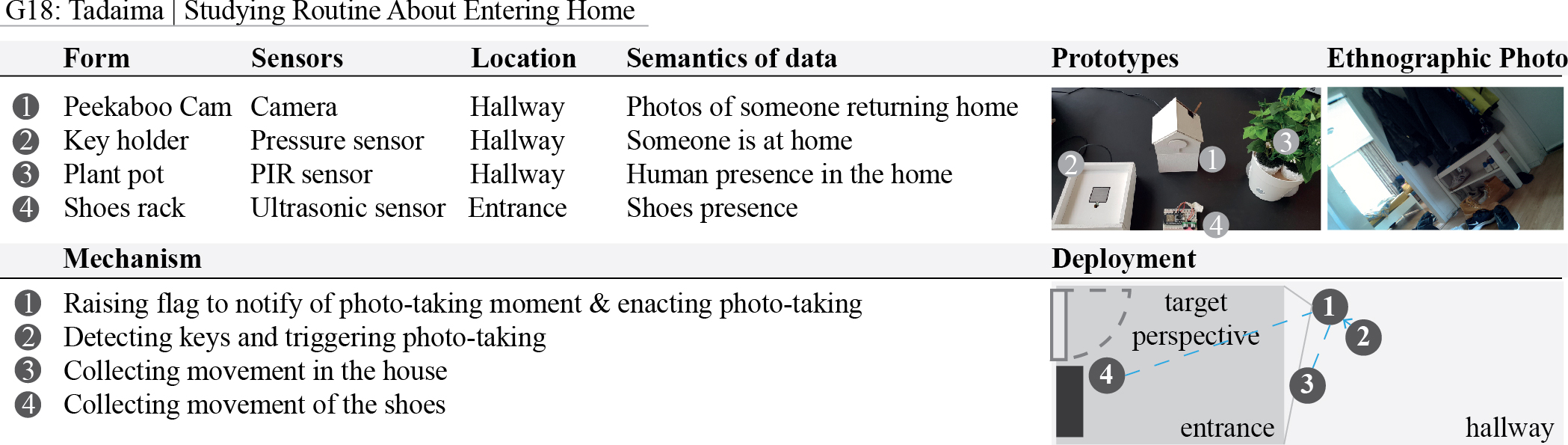}
  \caption{Team G18 designed a key holder that can naturally blend into a home entrance to trigger photo-taking when the keys are placed in their holder.}~\label{fig:project18}

\end{figure}

The Tadaima project by design team G18, aimed to study the activities that take place when someone is entering a home. In particular, G18 was interested in studying these activities through the changing patterns of clothing and shoes. They built a connected data-enabled system that targeted a clothing rack and a shoe rack. G18 found that the moment when participants placed their keys on a key holder tray when they returned home was the moment when clothing and shoe patterns could change. Therefore, G18 extended the key holder with an interactive sensor that could seamlessly blend data control into the home-entering routine. Due to the privacy concerns of their participants, G18 faced the challenge of capturing the clothing and shoe patterns without revealing participants' identities. As a result, they deployed the key holder to a location where the CPT camera would not capture the participants but it would capture photos of the changing clothing and shoe patterns without being blocked by the participants. 

In general, G18 built four connected data-enabled objects using CPT. These are listed as (1) to (4) in Figure~\ref{fig:project18}, which also shows how the objects were deployed in different locations and how they triggered different functions to prevent accidental photo-taking. G18 built a mechanism that would ensure \textcolor{minor}{the} CPT camera (1) only captured photos when participants placed their keys on the key holder (2). Additionally, after the participants placed the keys, CPT camera (1) raised a flag and played a sound to further ensure that the participants knew that the photo-taking had been triggered. G18 also introduced this data-control feature to their participants, showing them how they could always control the data-capturing mechanism simply by placing their keys on the holder. In addition, G18 built two other data-enabled objects, which are shown as (3) and (4) in Figure~\ref{fig:project18}. These two products assisted in capturing more data about the shoes and the human presence in the house without revealing the specific identification of any participants. With this system in place, G18 successfully captured multiple ethnographic photos of the changing patterns of clothing and shoes at the entrance of the home and matched the data with the photos through the data canvas (provided by CPT). The team G18 had an active discussion with their participants about the patterns of the clothing and shoes which showed different activities and habits. For example, the home participants shared their interpretations with the G18: ``Participant 1 went on a trip to another country during those 5 days, that's why the travel bag and the shoes don't show up until the next week.'' Additionally, the patterns of the changing shoes indicated the schedule of the participants: ``I \textcolor{minor}{(went)} to [university] and then to the gym, that's why you see the boots I take to class and the sports shoes I take when I got to the gym.'' Their privacy design for the data-enabled objects allow\textcolor{minor}{ed} them (\textcolor{minor}{T}eam G18) to capture rich home photos and facilitate their participants to co-interpret the photos together. Furthermore, the G18 shared that their participants were \textcolor{minor}{actually} excited to see their data-enabled objects as an intelligent system. 




\subsubsection{Peek-a-Boy: Studying Weekly Clothing Patterns in a Closet.}
\begin{figure}[h]
\centering
  \includegraphics[width=1\columnwidth]{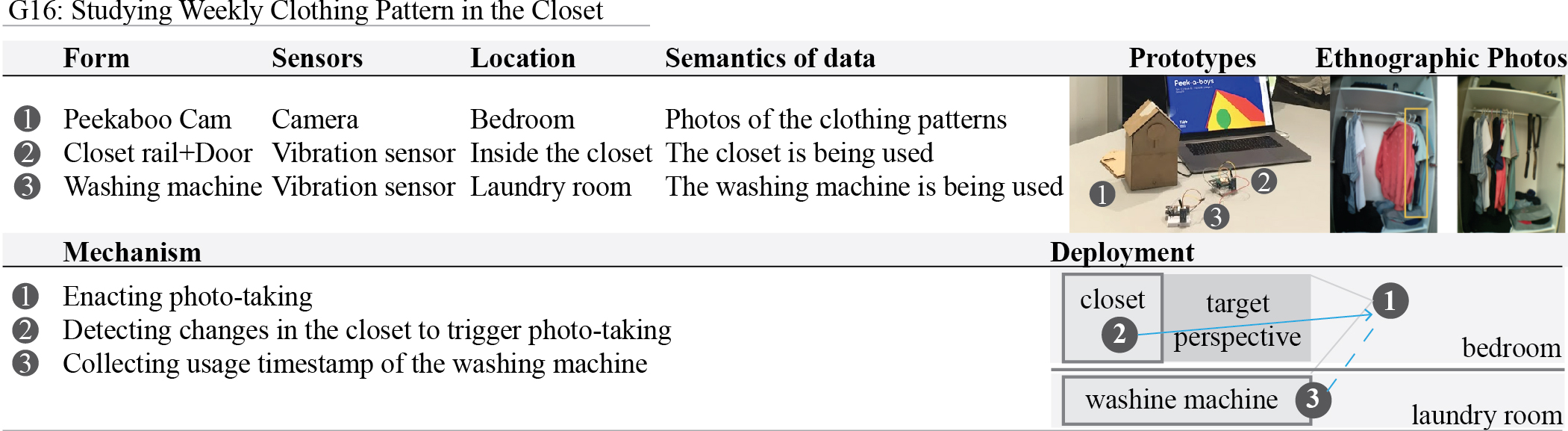}
  \caption{\textcolor{minor}{G16 implemented sensors into a closet and a washing machine to work with the CPT camera in capturing and analysing the relationship between clothing patterns in the closet and times when the washing machine was being used.}}~\label{fig:project16}

\end{figure}

The Peek-a-Boy project by G16, aimed at studying a participant's weekly clothing patterns in the closet. However, to capture such photos, G16 faced a critical challenge, due to the fact that their prototypes were deployed in what is usually a very private context, the bedroom. Furthermore, photos that would invade the participant's privacy, e.g., when changing clothes, could easily be captured in front of the closet. The team \textcolor{minor}{members} carefully discussed with their participant about the appropriate moment\textcolor{minor}{s} and observational perspective \textcolor{minor}{for} captur\textcolor{minor}{ing} the photo\textcolor{minor}{s}. As a result, they built three different data-enabled objects, shown as (1) to (3) in Figure~\ref{fig:project16}. G16 adjusted the observational perspective of the CPT camera (1) to only capture photos \textcolor{minor}{of} the closet. Additionally, they implemented sensors into the closet rod and door (2). Whenever the participant opened the door and removed the clothes from the rod, the camera notif\textcolor{minor}{ied} the participant and ask\textcolor{minor}{ed} \textcolor{minor}{them} to walk away so that the camera \textcolor{minor}{could} capture a photo inside the closet. To make their participant feel comfortable and avoid feeling anxious about such an invasion of privacy, G16 designed item (2) with an additional kill switch that the participant could press to postpone the photo-taking function anytime. Additionally, G16 \textcolor{minor}{was} interested in capturing more context about clothing, so they also designed a data-enabled washing machine (3) to capture its on and off data. Finally, \textcolor{minor}{T}eam G16 successfully deployed their data-enabled objects into their participant\textcolor{minor}{'s} bedroom and captured photos \textcolor{minor}{of} the clothing patterns in \textcolor{minor}{the} closet. \textcolor{minor}{T}eam G16 found that different arrangements in \textcolor{minor}{the} closet indicated their participant\textcolor{minor}{'s schedule}. For example, G16 interviewed their participants and found that \textcolor{minor}{they} arranged their weekend clothing and weekdays clothing into different sets\textcolor{minor}{, with ``t}he jacket and more fancy clothes [usually on the left]''\textcolor{minor}{, showing} how their participants ``want\textcolor{minor}{[ed]} to wear different [\textcolor{minor}{clothing on the weekend than on}] the [weekdays].'' At the same time, the team G16 also identified the patterns of the closet being full and empty by mapping the data \textcolor{minor}{related to} using the washing machine. However, G16 still found that their participants \textcolor{minor}{was} hesitant about interacting with the closet because the bedroom is the utmost private place\textcolor{minor}{,} where ``they felt a bit shy and reserved''. Nevertheless, \textcolor{minor}{the team's} privacy design (i.e., allowing the participant to always delete the photo, \textcolor{minor}{and maintaining an} observational perspective of no human presence) \textcolor{minor}{gave} their participants more confidence \textcolor{minor}{that} the data-enabled objects which would not accidentally capture other photos \textcolor{minor}{that} were not agree\textcolor{minor}{d to}: ``[the participant] felt confident \textcolor{minor}{(enough)} with the system that [\textcolor{minor}{they would}] not have to worry when the closet was closed.''
\begin{figure*}
  \centering
  \includegraphics[width=\linewidth]{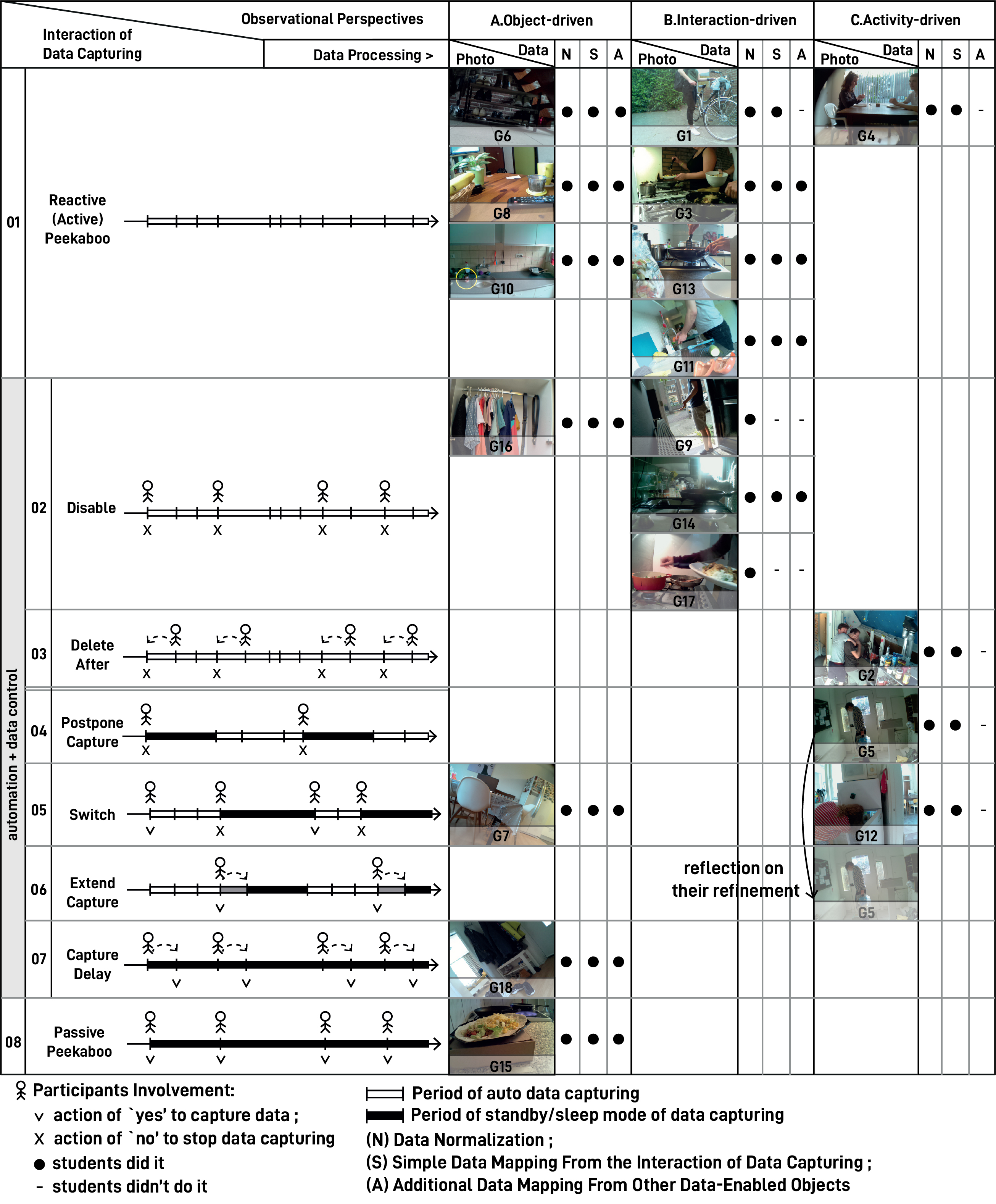}
  \caption{This table shows an overview of different output data (ethnographic photos made by their data-enabled objects) by presenting different observational perspectives from \textcolor{minor}{the} 18 design projects. We also presented their combinational strategies \textcolor{minor}{for} designing mechanism \textcolor{minor}{for} data capturing\textcolor{minor}{,} from active mode to passive mode, and data processing (N: data normalization; S: simple data mapping; A: additional data mapping from other data-enabled objects). 
}~\label{fig:projects}
\end{figure*}
\subsubsection{Summary.}
The two projects demonstrated how \textcolor{minor}{the} CPT enabled the teams to build data-enabled objects specific for particular homes and studies. The CPT provided a scaffolding framework so that every teams could actually test their design without interfering with their participants' privacy needs.
All student design teams were engaged in design explorations enabled by \textcolor{minor}{the} CPT, which allowed them to explore and share many details related to privacy concerns. Their design projects enabled us to further analyze the design aspects which are presented in the following sections. These design aspects present collective examples and rationales from the 18 cases, encompassing the designing, implementing, and testing phases. The following presents the five design aspects (form, notification of data capturing, observational perspective, interaction \textcolor{minor}{mode} of data capturing, data processing \& mapping) we identified from the 18 cases. \textcolor{minor}{An overview of how 18 design projects used the different design aspects is presented in Figure~\ref{fig:projects}.}

\subsection{Form}
The design aspect of Form is about designing for familiarity and how well data-enabled objects blend in with everyday experience, using two specific instances: the estranged and the everyday form. To fit their data-enabled objects to a target context and ethnographic style, the student design teams designed \textcolor{minor}{their objects using either} the everyday form\textcolor{minor}{, which blends in,} or the estranged form\textcolor{minor}{, which stands} out and attract\textcolor{minor}{s} attention. Shaping the products towards an invisible, blending-in form may capture rich in-situ data yet raise participants' uncomfortable feelings of being watched without notice. This was left as an open challenge for the student design teams to address, whether in the process of making their objects or in reflecting on them—not only as an educational experience, but also to enrich our study results. We present instances of the two styles of Form in the following sections.

\subsubsection{Everyday Form}
\begin{figure}[h]
\centering
  \includegraphics[width=1\columnwidth]{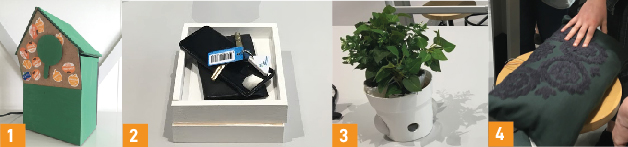}
  \caption{Design Cases of Everyday Form from \textcolor{minor}{the d}esign \textcolor{minor}{t}eams. 1) G14 decorated their data-enabled objects with green paints and stickers to match the home style of their participants. 2) G18 designed a key holder that \textcolor{minor}{could} be naturally blend into home entrance to trigger photo-taking when \textcolor{minor}{keys were placed on it}. 3) G18 integrated \textcolor{minor}{a} PIR sensor into a `flower pot' to detect human presence.  4) G4 designed a cushion \textcolor{minor}{that could} detect the pressure of a person \textcolor{minor}{sitting on it}.}~\label{fig:everyday}
\end{figure}

Everyday form was the most frequently used form by the student design teams. This form blends well into the environment (see Figure~\ref{fig:everyday}), and has a peripheral existence~\cite{bakker_2015} that possibly connects to the participants' daily routine. Participants can be aware of it and interact with it naturally. As design team G18 explained, ``The system should be integrated into the home [...] respectfully so that participants do not need to feel that they are constantly watched.'' For example, G3 painted their data-enabled objects to match the color of the home (see Figure~\ref{fig:everyday}-1); G18 blended a trigger function with a key holder (see Figure~\ref{fig:everyday}-2), and designed sensors into a plant pot (see Figure~\ref{fig:everyday}-3); G4 integrated pressure sensors into a cushion to capture the period of time someone spent sitting on it (see Figure~\ref{fig:everyday}-5). After the studies, the participants confirmed that the teams' designs had integrated the data-enabled objects well with the users' daily routines. However, such seamless integration into a home can also bring about privacy issues, as \textcolor{minor}{one of} G17's participants commented, ``This gave me a somewhat unsafe feeling and this was a small privacy issue.'' Everyday form can be efficiently used to capture natural behaviors by being blended into the environment. 


\begin{figure}[h]
\centering
  \includegraphics[width=1\columnwidth]{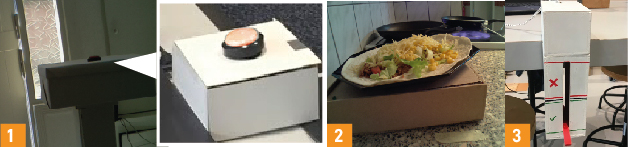}
  \caption{Design Cases of Estranged Form made by the student design teams. 1) G5 built a `red button' as a metaphor \textcolor{minor}{for} `emergency use' to stop data capturing. 2) G15 built a special stage for participants to make food blog photos after cooking, but their form did not convey \textcolor{minor}{a} strong trigger to their participants. 3) G12 designed a big slider for participants to \textcolor{minor}{use to} control the data capturing. }~\label{fig:estranged}
  \vspace{-7mm}
\end{figure}
\subsubsection{Estranged Form}
We found that some teams used an estranged form, the opposite of the everyday form: These data-enabled objects and sensing objects do not fit the environment naturally, but still can be useful for capturing special aspects of social interaction. This was the case with teams G5, G15, and G12, who built some of their auxiliary sensing objects with an estranged form. In designing their data-enabled objects so that they would not seamlessly blend in with the environment, they aimed to provide a distinctive feedforward~\cite{djajadiningrat_2002}. These objects were meant as probes for additional interactions that were not fully connected to the participants' daily practice\textcolor{minor}{s} or routines. After their studies, G15 and G5 realized that they had captured fewer photos than they had expected. G5 found their form of a red button (see Figure~\ref{fig:estranged}-1) surprisingly did not inspire emergency use for blocking data, but rather provided a game-like feeling. Their participants felt so curious that they kept pressing the button. As a result, the data-enabled objects in G5 captured fewer photos than expected. On the other hand, G15 built a special stage (see Figure~\ref{fig:estranged}-2) for participants to use for taking food blog photos after cooking. However, their participants sometimes forgot to do so due to \textcolor{minor}{the} fact that the form of the special stage was a simple, plain box, which did not convey a strong trigger to the inhabitants despite not fitting into the environment. In these cases, the estranged forms clearly did have a novelty effect and evoked unusual behavior initially, thus suggesting that such forms could provide a novelty to trigger social interaction for ethnographic needs. 


\subsection{Notification of Data Capturing}
Notification of data capturing is a design aspect \textcolor{minor}{that involves} exploring the expressions \textcolor{minor}{used} to notify participants that data was being captured. To better capture natural interactions in the home, the teams integrated their data-enabled objects with the environment. Such integration can bring \textcolor{minor}{the} anxiety of being watched \textcolor{minor}{by} a black box\textcolor{minor}{, with the} participants hav\textcolor{minor}{ing} no agency in controlling the data collection. Therefore, designing a means of notification before data capturing prevents the data-enabled object from becoming merely a black box\textcolor{minor}{, instead allowing the} participants to consider the possible outcomes and harms of having their data recorded at any particular time, thus allowing them to make better choices for themselves on blocking or approving the data collection. As one of the teams explained, ``Inform(ing) the user about the possible outcome is also a very important approach. It can [also allow] the user [to evaluate whether] when the prototype is malfunctioning.'' (G11) Designing notification during data capturing allows participants to be prepared for the capturing. The team recorded that one of the participants acknowledged this advantage: ``Because of the fact that he heard the Peek-A-Boo making a photo, it became clear to him that moments of privacy in the household would be randomly captured.'' Even by designing a simple expression, such as a notification light, these small changes can improve the process of data collection, as one team reported: ``Adding a few LEDs and a stop button made them feel in charge, while it did not really impact the quality of the data. It is important to always keep in mind that you are designing for humans and that small changes like this can make a doubtful idea (filming in someone's home) a successful user study both for the participants and the research team'' (G13). However, a challenge arises in determining how many notifications the data-enabled objects should provide. Too many notifications can interfere with ethnographic practice, for example by creating too much noise, while too little notification can make the data-enabled objects too much like a black box. 

\begin{figure*}
  \centering
  \includegraphics[width=\linewidth]{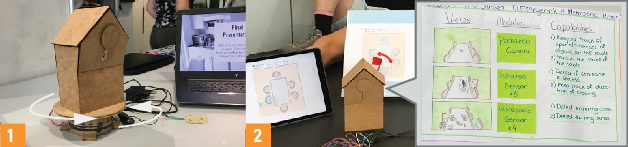}
  \caption{Notification of Data Capturing. 1) \textcolor{minor}{T}eam (G11) built a turning table for the camera to exaggerate the expression of `turning away' when the camera was disabled. 2) \textcolor{minor}{T}eam (G7) designed \textcolor{minor}{a} flag-waving \textcolor{minor}{mechanism} to convey a message of `I sense you now' when the system detecting a human participant nearby. }~\label{fig:notification}
\end{figure*}

\subsubsection{Turning Away (Notification of Photo Taking Only)}
Besides using the notification expressions provided by the default CPT (e.g., an interactive enclosure, a flag, and a notification sound), the teams also built additional expressions to exaggerate these notifications. Team G11 built a turntable for the camera to exaggerate the expression of turning away when the camera was disabled (see Figure~\ref{fig:notification}-1). One of the members of this team highlighted the effectiveness of this design: ``I noticed that it was appreciated when the entire [CPT camera] could be turned away from the general view. It provided a real sense of privacy because it really felt like the system was looking away'' (G11). The turntable also made their participants see the data-enabled objects as a human-like agent: ``The addition of the turning [CPT camera] proved to be very interesting, since it provided the system with a more human feeling, which can influence participants' views on privacy on a more social level'' (G11). This example also shows another important aspect that \textcolor{minor}{was} widely discussed by every student design teams: \textcolor{minor}{w}hether to capture people in the photo, and what the appropriate observational perspectives of the camera should be (section 6.4). 

\subsubsection{I Sense You Now (Notification of Data Sensing \& Photo-Taking)}
Team G7 added a flag-waving design to convey the message ``I sense you now'' whenever the system detected participants nearby. As described by one of the team members: ``If the system detects that there are humans entering the area, the flag will be raised. So in this way, people know that when the flag isn't raised, the system did not recognize their presence'' (G7). This ensured that their participants were aware not only of the data-collection moment (i.e., the photo-taking moment) but also the sensing moment (see Figure~\ref{fig:notification}-2). \textcolor{minor}{This example shows a wider design space for the notification of data capturing in which the notification is not only for photo-capturing but also for capturing sensor data.}

\subsection{Observational Perspectives}
Observational perspectives determine where exactly data-enabled objects are deployed and what angles they can capture. Because the provided CPT includes a camera, the challenge of \textit{whether to capture participants in the photo} became an important concern for some design teams as it relates to balancing privacy needs with research needs (G1, G2, G5, G8, G12, G14, G17). For example, directly capturing human behaviors and facial expressions can tell rich stories about life in a home, yet it also invades participants' privacy to a high degree. the student design teams repeatedly mentioned that ``the right angle'' for taking photos needed to be found. We identified three main angles, or perspectives, that were taken by their data-enabled objects: object-only, interaction-driven, and activity-driven. It should be noted that in the case of a limited population such as a household, not capturing faces would not represent a sufficient level of privacy protection—at least to anyone familiar with the inhabitants of the home. In this sense, privacy protection from the three perspectives is more a matter of degree than absolute. With this framework, we will focus on the nuances of the teams' privacy-aware designs as well as their experiences and reflections on working through such a design research process. The relevant photos can be seen in Figure~\ref{fig:projects}. 

\subsubsection{Object-Only Perspective (Object)} 
Design teams G6, G7, G8, G10, G11, G15, and G18 decided to capture \textcolor{minor}{their} chosen context from an object-only perspective, meaning one that captured only objects as a means for studying their participants' home routines. This was based on the visual features of the objects and their displacement over time. The rationale here was to reduce any feeling of ``being watched'' for the participants (G7), and to prevent participants from acting abnormally when becoming aware of a camera facing them (G16). The teams were also interested in how many stories could be found by only capturing objects in a home; therefore, they added more data-enabled objects with different sensors and interactions, all aimed at preventing pictures being taken of any participants. For example, Team G18 made \textcolor{minor}{a} key holder and a shoe rack that could detect motions nearby and only captured photos when participants placed their keys on top of the holder and walked out of the room. Observing these object traces can tell stories of a home or a person, for example, they can tell which style of clothing people usually choose for different occasions (G11), or how they order cooking ingredients (G10). Furthermore, the object-only perspectives also captured social interactions \textcolor{minor}{with the} participants. Some participants even put special objects in front of the data-enabled objects to be photo-captured (G3). Although specific activities cannot immediately be recognized from these object-only photos, the photos pointed the student design teams in the right direction for looking into details of sensor data or even for extending the data-enabled objects to capture new data to interpret contexts, activities, preferences, and lifestyles beyond what the photos presented. As Team G15 stated, ``we can analyze meals, meal times, meal sizes, and also predict habits [if] we extend this over a time period.'' Most of the participant feedback on the object-only perspective was positive, and the participants expressed few concerns about data privacy. However, Team G15 mentioned in their reflections that issues of privacy did not completely disappear. The object-only perspective can still be intrusive because objects could still imply sensitive personal information (e.g., sexual preferences or personal health): ``mainly people cannot be in the picture, but also other privacy sensitive objects need to be out of the picture'' (G17).

\begin{figure}[h]
\centering
  \includegraphics[width=1\columnwidth]{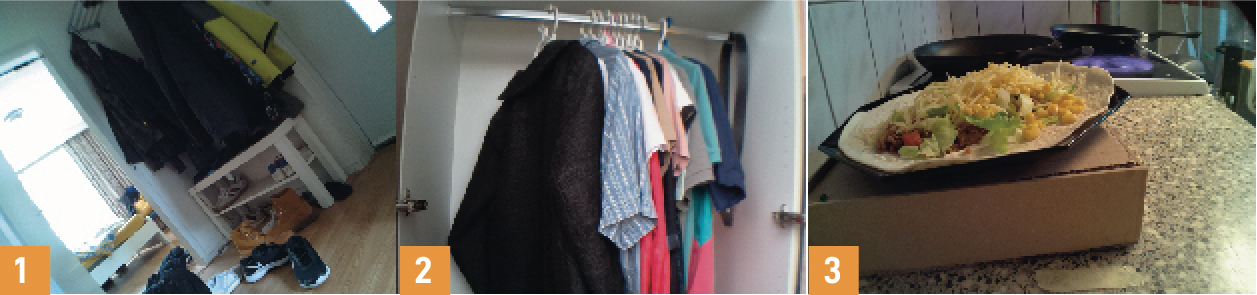}
  \caption{Photos indirectly showing human presence (i.e., through tracing objects) taken from the student design teams' projects.}~\label{fig:indirect}
  \vspace{-5mm}
\end{figure}

\subsubsection{Interaction-Driven Perspective (Object + Human Body)}
\begin{figure}[h]
\centering
  \includegraphics[width=1\columnwidth]{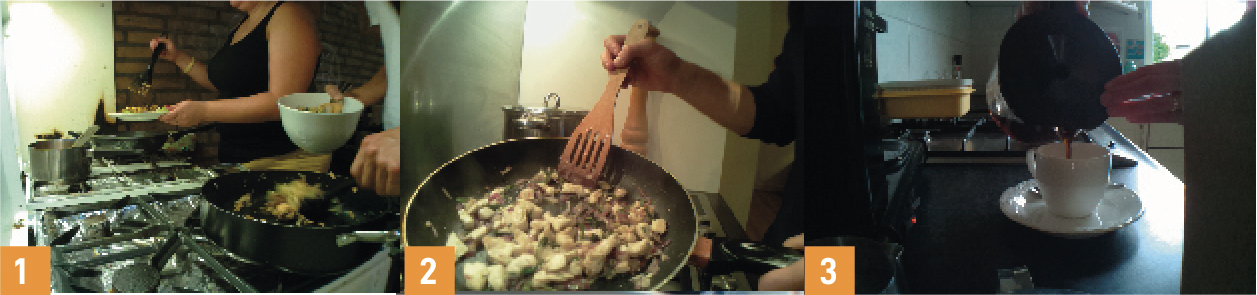}
  \caption{Photos semi-directly showing human presence (i.e., bodies\textcolor{minor}{and} hands) taken \textcolor{minor}{from} the student design teams' projects.}~\label{fig:semidirect}
  \vspace{-5mm}
\end{figure}

Design teams G1, G3, G9, G11, and G17 decided to take a perspective in which they focused on the interaction between participants and objects. This interaction-driven perspective captured objects and participants, but only the bodies of the participants, not their faces. These teams provided a similar rationale as those using the object-only perspective: They were concerned that ``most people did not want to be recognized in the pictures'' (G1). These teams adjusted the position of their data-enabled objects or designed some form of visual blocking to ensure that the objects would not capture faces (see Figure~\ref{fig:semidirect}). Team G14 argued that by only focusing on the bodies of their participants, and blocking the view towards their faces, they prevented the recording of their personal identities. Meanwhile, they could still capture activities and extract meaning from recorded body postures. After the studies, these teams found they had captured photos that successfully presented in-situ interactions between participants and daily objects, such as the ways that the participants cooked their food (G3, G11, G13, G17). However, although no faces were recorded and the photos were largely anonymous, some participants still had the feeling that they were being watched. This resulted in having some effect on their behavior. Team G14 illustrated, for example, that ``[the] influence our system had on the house was...they [the participants] did not walk around with little clothing, because they might be photographed this way.''

\subsubsection{Activity-Driven Perspective (Object + Human)}
\begin{figure}[h]
\centering
  \includegraphics[width=1\columnwidth]{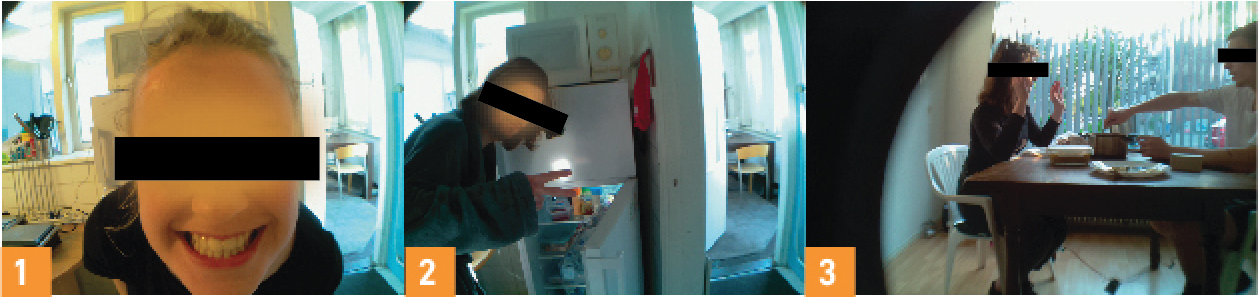}
  \caption{Photos showing human presence directly and identifiably as taken by \textcolor{minor}{the} design teams' data-enabled objects, \textcolor{minor}{anonymized} for publication.}~\label{fig:direct}
  \vspace{-5mm}
\end{figure}

Design teams G2, G4, G5, and G12 decided on an activity-driven perspective, capturing entire scenes, including objects and participants, with the approval of the participants, to study social activities in the home. They were also concerned about invasive photos, but instead of adjusting the photo-capturing perspective, they relied on interactive controls (see interaction \textcolor{minor}{mode} of data capturing in Figure~\ref{fig:projects}) with which participants could determine accessible and inaccessible photos. After the study, they all \textcolor{minor}{had} captured series of photos that directly showed activities and the presence of people (see Figure~\ref{fig:direct}); the participants of Team G1 even posed in front of the camera. These design teams found that their participants had ``no privacy concerns'' because they understood the clear purpose of the study (i.e., ``for research only'') (G5) or because they were able to control whether the photos were taken (G4). In fact, some of the student design teams, in reflecting on this\textit{ no concerns} response, were surprised at how fast their participants got used to the camera (G9). On the other hand, although the participants approved of their faces being clearly captured before the study, some of them changed their minds during the debriefing and asked that their faces be blocked. In reflecting on the design choice between interaction-driven and activity-driven perspectives, \textcolor{minor}{T}eam G1 noted: ``...in the end it is better to be careful, as even these people asked for their faces to be blurred out afterwards. I think they might not have remembered at that moment that these pictures could be used for the research as well.''

\subsection{Interaction \textcolor{minor}{Mode} of Data Capturing }
To study different situations in the home, the student design teams extended the abilities of their data-enabled objects to perform more flexible data capturing by adding sensors and actuators. For example, an extension could be a fully automatic privacy-aware system triggered by sensors or a fully manual control system to be triggered by the participants. To determine what kind of extensions they needed, the groups had to decide how much participant involvement they would like to see or allow for. This was, again, a trade-off: On the one hand, adding some form of data control for participants might allow them to feel more in control. On the other hand, it could also change behaviors and thus not match ethnographic needs. In the following section, we present eight different interaction modes for data capturing built by the student design teams. \textcolor{minor}{An overview of the right different interaction modes can be seen in Figure~\ref{fig:projects}.}

\subsubsection{Reactive (automation) Mode}
The reactive mode allowed for fully automatic data-capturing triggered by sensors, with no additional interactive controls for use by the participants. This mode was used by teams G1, G3, G4, G8, G10, G11, and G13. By connecting to specifically designed auxiliary sensing objects, the data-enabled objects could detect changes in the environment and enact data capturing accordingly. For example, G18 only captured photos when detecting someone sitting on the living room couch. They built sensors to activate the data-enabled objects and capture data in the right moment. However, not all of the inhabitants in the homes were comfortable with this interaction mode. G1 mentioned that privacy intrusion was not only attributed to the camera, but also to the auxiliary sensing objects they deployed: ``All residents were aware of the deployment of sensors in the toilet [room], which could have made them aware [of] the duration inside the toilet [room]''(G1). Design team G4 reflected that ``[in] the end, [we] realize that this is a significant lack of respect from the system and would have discouraged and made the users feel uncomfortable.''

\subsubsection{Disable Mode}
The disable mode was an interactive control that participants could use to stop data capturing immediately. This mode was used by teams G6, G9, G14, G16, and G17. Team G9 built a disable button that participants could press. Teams G14, G16, and G17 allowed participants to wave their hands at the data-enabled objects to signal \textit{No}. G6 turned the participants' doorbell into a disable button; when someone pressed the doorbell, the data-enabled objects stopped taking photos for a few minutes. This team wrote that ``this would [protect] the privacy concern of taking pictures of people coming in and household members opening the door'' (G6).

\subsubsection{Delete-After Mode}
The delete-after mode was an interactive control that participants could use to delete data after it has been captured by a data-enabled object. This mode was proposed by \textcolor{minor}{T}eam G2 because they found that participants would not always notice the precise moment when a photo was taken. A mechanism for deleting ``after the fact'' (G2) prevented invasive or otherwise unacceptable photos from being stored, thus inspiring trust in the participants.

\subsubsection{Postpone-Capture Mode}
The postpone-capture mode was an interactive control that participants could use to postpone a data-capturing moment. Design team G5 found this mode useful for their participants, who used it to protect their more intimate interactions with others, such as kissing or hugging.


\subsubsection{Switch Mode}
The switch mode was a control that participants could use to flexibly decide when to stop and start data capturing. It was used by teams G7 and G12. Teams G7 designed a button for disabling data collection immediately at any time. G12 designed a slider that allowed participants to turn the data-enabled objects on or off without the use of any automation or timer.

\subsubsection{Extend-Capture Mode}
The extend-capture mode was proposed by team G5 as an interactive control that participants could use to ask the data-enabled object to extend its data-capturing period. G5 found from the earlier reactions of their participants that they were having fun with the the extend-capture mode, and so they conceptualized another mode which could be  combined with a fun form (which \textcolor{minor}{could be considered} one of the estranged form\textcolor{minor}{s, meant} to attract participants' attentions) to encourage participants to trigger more data capturing.

\subsubsection{Capture-After Mode}
The capture-after mode was an interactive control that participants could use to trigger data capturing after a particular moment. This mode was used by team G18, who found that the delay in photo-taking allowed for participants to prepare for the photo and to decide how they would like to be captured. In fact, their participants sometimes posed just for the data-enabled objects to capture them (see example in Figure~\ref{fig:projects}-G18). 

\subsubsection{Manual Mode}
The manual mode was an interactive control that participants could use to instantly trigger data capturing. This mode was used by team G15, who designed a mechanism called ``Picture Your Dinner.'' They found, however, that this mode required  a strong motivation or incentive for participants to use it. G15 reported that their participants remembered to capture their food in the beginning of the research, but even then they forget to do so reliably.

These eight patterns presented a spectrum \textcolor{minor}{for} designing the interaction \textcolor{minor}{mode} of data capturing for data-enabled objects from active agents to passive agents, or \textcolor{minor}{for} designing the human data intervention from passive control to active control. Depending on different situation\textcolor{minor}{s}, the data-enabled objects required different capabilities \textcolor{minor}{for} captur\textcolor{minor}{ing} data\textcolor{minor}{.} 

\subsection{Data Processing \& Mapping}
To support participants and researchers in efficiently making sense of data during the debriefing and interviews, the teams adjusted the sensor values in order to allow for better mapping of the data with semantics when visualizing it on the Data Canvas. In the beginning, they faced some challenges in dealing with their data and photos. As \textcolor{minor}{T}eam G7 admitted, ``We were really confused on how to get started and actually make sense of the data.'' The sensor values were set by different benchmarks, thus some sensor data presented in an exaggerated changing peak whereas some was captured in a narrow peak. As a result, when mapping all of the data with exaggerated peaks, the Data Canvas presented difficulties for the teams and their participants in analyzing the differences and commonalities among the data results (see Figure~\ref{fig:groupdata}). The teams used mainly three ways to improve the legibility of the data from the data-enabled objects. Such improvement of data legibility also allow\textcolor{minor}{ed} participants to make sense of the captured data about themselves. The three ways are presented as follows:

\subsubsection{(N) Data Normalization for Every Data-Enabled Object}
\begin{figure}[h]
\centering
  \includegraphics[width=1\columnwidth]{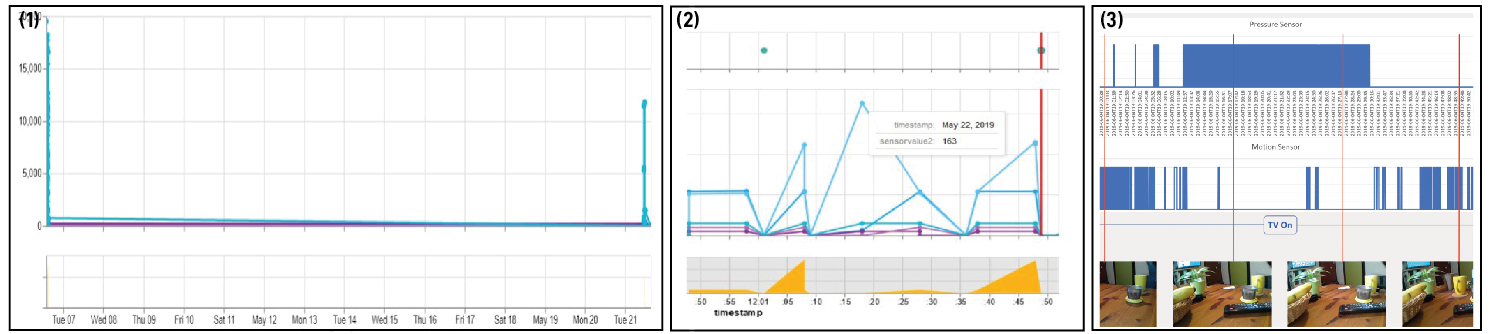}
  \caption{1) Scaling issues in the beginning of the prototyping phase. \textcolor{minor}{Team }G1 found that ``The data in the Excel-sheet is recorded correctly and the [data-enabled object] reacts to the commands. Unfortunately, the data is not shown correctly on the local graph. Fortunately, we recently discovered it might be because a value of 1 is too small to notice.'' 2) Proper scaling shown in the Data Canvas from \textcolor{minor}{Team} G7. They adjusted and normalized sensor values for a better presentation on the Data Canvas. 3) Further data mapping by \textcolor{minor}{Team} G8 with photos for further interpretation.}~\label{fig:groupdata}
\end{figure}

Instead of cleaning the data after data collection, all of the teams adjusted the sensitivity of the sensors for every data-enabled object so that all data could be automatically visualized on \textcolor{minor}{the} Data Canvas during the data collection (see Figure~\ref{fig:groupdata}-2). The teams also further discussed what possible semantics their data-enabled objects could generate and \textcolor{minor}{how} to define whether their products should be designed to capture digital or analog data. For example, Figure~\ref{fig:groupdata}-3 shows how \textcolor{minor}{T}eam G8 defined their data-enabled objects to capture usage of the TV and the couch as digital data that only shows in-use or not in-use information.

With the help of a normalization step, the student design teams could better link semantics to the data and thus make better sense of it. For example, they were able to identify interesting moments to interview their participants about. The participants also appreciated the visualizations. They reported a better understanding of the data using the \textcolor{minor}{D}ata \textcolor{minor}{C}anvas during the debriefing to evaluate the data. While the data mapping could inspire the student design teams to explore meaningful patterns\textcolor{minor}{,} the resulting challenge was in determining how much data should be used and to avoid accidental correlations between different data streams that would invade privacy more than previously agreed upon with the participants and designed for. 
\subsubsection{(S) Simple Mapping through the Interaction \textcolor{minor}{Mode} of Data Capturing}
Every piece of data and photo presented on the Data Canvas \textcolor{minor}{was} conceptually annotated by simple semantics determined by the data-control mechanisms. Since the teams built data-control mechanisms that allowed the participants to evaluate every situation and decide whether to enable or disable data collection, every final captured photo showed the availability of a home to be captured. At the same time, any time that the data collection was manually stopped annotated a moment when the home was unavailable to be captured. More specifically, every intervention that a data-enabled object or participant made was automatically annotated with simple semantics that told about the situation in the specific moment. For example, \textcolor{minor}{T}eam G2's delete-after data-capturing mode allowed them to know when the photos were deleted, providing an anchor to be discussed during the interviews. The team discussed with the participants why and how they felt unavailable. Other modes, such as postpone capture, capture after, and manual mode, also allowed the teams to ensure every captured photo was consented to by their participants and at least represented \textit{an available moment} or \textit{in the activity} behind each photo.


\subsubsection{(A) Additional Mapping of Data-Enabled Objects Distributed in Places} 
\begin{figure}[h]
\centering
  \includegraphics[width=1\columnwidth]{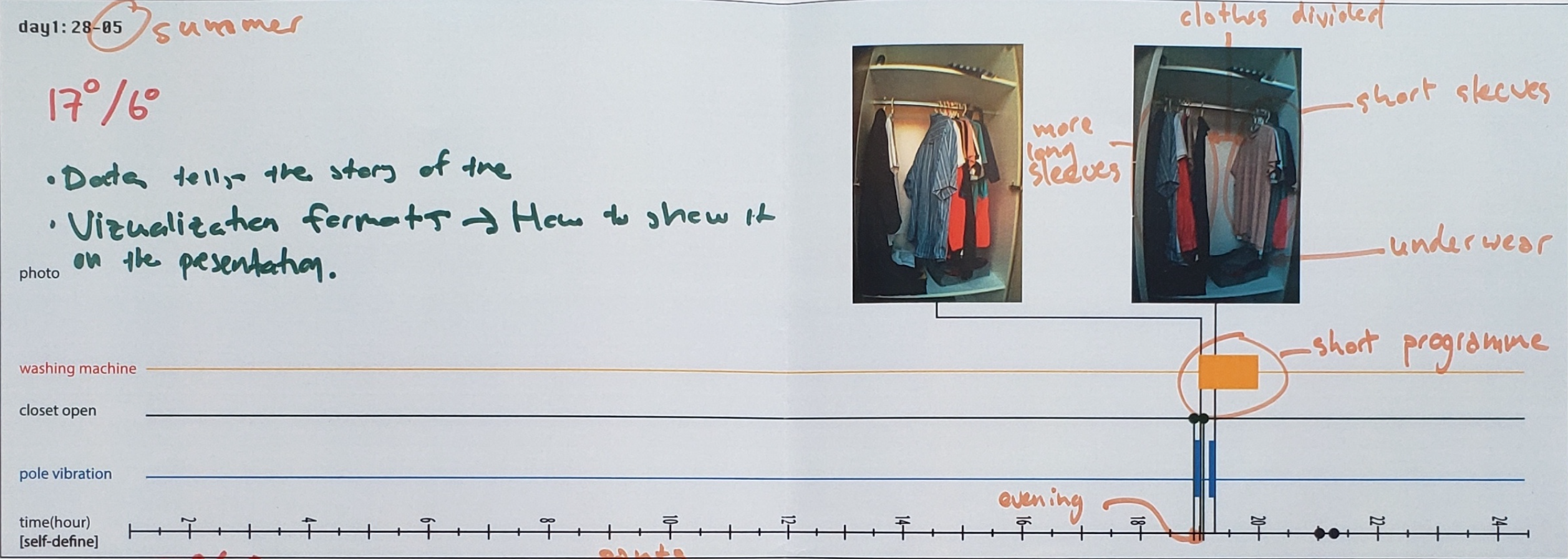}
  \caption{\textcolor{minor}{G16 mapped the photos of a closet with activation timestamps of the household washing machine to further interpret the collective laundry routines.}}
  ~\label{fig:G16annotation}

\end{figure}

Besides the annotation from the data-control mechanisms, the teams G3, G6, G7, G8, G10, G11, G13, G14, G15, G16, and G18 mapped additional semantics from their sensing objects distributed in multiple places. This additional data mapping enriched the teams' understanding of the context by also capturing different angles or sub-contexts than the photo perspective. For example, G8 mapped photos of the coffee table with activation timestamps of TV and couch use. They identified that, sometimes, the TV was turned on without anyone sitting in front of it. Further investigation revealed that the TV was switched on purely for background noise (see Figure~\ref{fig:groupdata}-3). G16 mapped the photos of a closet with activation timestamps of the household washing machine to further interpret the collective laundry routines \textcolor{minor}{(see Figure~\ref{fig:G16annotation})}. This additional data mapping from other types of data-enabled objects supported the teams not only in understanding multiple contexts but also in interpreting data into a temporal narrative such as an activity flow.

\subsection{Summary}
The 18 design projects exhibited common and diverging points in rationale and design choices. Apart from the information in the team design reports, we analyzed each team member's reflections to get an understanding of how individual team members made decisions and formed appropriate reactions to privacy issues in design research. For example, we found that \textcolor{minor}{T}eam G4 came to understand the importance of allowing participants to take part in data control; G17 realized that every new tactic they came up with would have new implications for privacy. The analysis of \textcolor{minor}{the} student design teams' design reports allowed us to derive five design aspects for designing data-enabled objects concerning its \textit{form, observational perspective, notification of data capturing, interaction \textcolor{minor}{mode} of data capturing and data processing}.

\section{Feedback on the Toolkit}
The previous sections give a detailed account of design rationales and outcomes from 18 design cases. From these cases, we see that our CPT successfully enabled multiple design explorations for and with data-enabled objects in the wild. Five design aspects were identified and presented. In this section, we \textcolor{minor}{discuss} the teams' feedback on the CPT and on using it in their design research.

\subsection{Creating Open Design Templates for Data-enabled Objects }
The design teams shared that the CPT provided them with a template which the ``researcher can follow as a standard [which is] useful so that everyone is on the same page when it comes to privacy'' (G16). The template allowed them to begin with a similar type of data-enabled object, the data-enabled peekaboo camera in the CPT, and explore how to fit it to the type of study context they \textcolor{minor}{were} target\textcolor{minor}{ing}. The template provides design researchers choices that \textcolor{minor}{gear} them \textcolor{minor}{in the right direction for} design\textcolor{minor}{ing} privacy-aware data-enabled objects to the right direction. 

The programmability provided by the CPT allowed the teams to modify the expression of the camera, for example by adding a raising flag, a notification sound, or \textcolor{minor}{an} interactive shutter. These small features helped the teams \textcolor{minor}{to} notice the importance of designing the notification and feedback functions for data-enabled objects and to extend this into building other types of data-enabled objects. One of the members from G13 highlighted that ``the most interesting lesson I learned...is that small changes in a design can have a big impact on how safe and respected the participants of the user study feel'' because ``something as simple as a flag that raises up when taking the photo received a lot of positive feedback.'' G7 extended the expression design to better inform their participants not only about the peekaboo camera but also other types of data-enabled objects that were sensing them.

Additionally, \textcolor{minor}{the} CPT provided an open design \textcolor{minor}{for} the interaction with data collection. For example, the teams \textcolor{minor}{could} design their own button or other sensing mechanism for triggering the data collection. 
Some of the teams designed mechanisms that called for participants to take an active role in setting off the data collection. For example, team G13 reflected on the importance of designing agency for their participants: ``Being able to give the residents an opportunity to disable the camera with [our] button, not only prevents possible privacy concerns, but more importantly it give\textcolor{minor}{s} insights in\textcolor{minor}{to} how often people are feeling uncomfortable having a camera around'' (G13). However, not all of the participants took an active role in working with the data-enabled objects. Because of different levels of participant reactions, the teams built different types of data control mechanisms, which further allowed us to unpack a spectrum ranging from automatic to manual data capturing.

Finally, the teams appreciated the built-in connectivity and data-sharing interfaces enabled by the CPT. These two features provided a template that the teams could use to work with data locally and to debrief their participants. The teams relied on this template as a fundamental data infrastructure that allowed them to safely build and conduct various kinds of data collection within a household, and to do so while taking privacy into consideration. One member of the team G5 mentioned, ``the fact that the participants knew that the data would only store data locally and \textcolor{minor}{[that the CPT]} is not connected to the internet helps a lot with the privacy concerns in my opinion.'' One of the members from G3 added this thought on dealing with privacy issues:  ``I also learned the proper procedure of making consent forms and \textcolor{minor}{[the]} process \textcolor{minor}{[of asking]} for privacy permission with users [as a way] to have complete respect \textcolor{minor}{[for]} users and \textcolor{minor}{[a]} professional way to collect data.''

In these examples, the CPT allow\textcolor{minor}{ed} the team members to implement and test their design hypothesis for privacy-aware data-enabled objects and deploy the objects into the studied context. Even though the student design teams might have been inexperienced in designing and researching with data-enabled objects, the CPT helped \textcolor{minor}{them} reflect during their design practice and make their own privacy design choices. For example, the teams were able to learn from their participants and to explore the essential qualities \textcolor{minor}{needed} for data-enabled objects. One member of G16 said ``... before now I had never gone into detail or designed taking into account this aspect (privacy), although I know the importance that this has and will have in a world where devices will be increasingly intelligent and communicating with each other.'' 


\subsection{\textcolor{minor}{Foregrounding} Privacy to be Investigated in the Wild}
By using the CPT, the design teams were pushed to think about and design for privacy. 
The design teams independently found that the CPT provided them with a controversial design challenge of deploying a privacy-sensitive camera into a private home. As cameras have become commonly known to easily capture, share, and deliver rich but potentially intimate information, people can be very reluctant to install any camera in their home. A member of \textcolor{minor}{T}eam G1 described the situation thus: ``We had to think of privacy while incorporating a camera in our data gathering process, [and it] has been a challenge, because just by simply mentioning the notion of \textit{camera} to someone it immediately cancels the value of privacy'' (G1). This challenge provided the teams a situation in which they had to foreground their participants' privacy concerns and critically observe the line between invasion and privacy. ``The fact that we had to tackle the issue of privacy, but using at the same time a camera in our design made it even harder, and honestly it did not make much sense to me in the beginning, but after giving it a second thought I discovered that this situation is ideal and it fits perfectly the real-life situation. This fact, the usage of the Peekaboo camera, provided me with the perfect situation of observing and feeling how easily the line between invasive and private can be crossed, especially while designing'' (G1). As such, the design teams were motivated to more deeply observe and consider the privacy requirements of their participants. As noted by a member of \textcolor{minor}{T}eam G11: ``[We] got to experience the annoyances and the benefits of the system. It also opened my eyes about just how intrusive such a system really is. Privacy has not been something that [we] have had to deal with in previous designs'' (G11).

The design teams learned lessons from their design actions that led them to reflect on better ways to deal with data in their future work, as suggested in a comment by a G16 member: ``By knowing how to handle this type of data, as well as how to prevent it from happening, allowed my team and [me] to prioritize the opinions of the user since we did not want him to [have to go] through any hardships'' (G16). Additionally, team members reflected on diverse privacy preferences: ``I always thought that nobody wanted that [camera], but now I know that it can be different, through the way it is used and where and if the participants are [OK] with it taking pictures'' (G1). Members of \textcolor{minor}{T}eams G1 and G5 even changed their own opinions toward accepting a camera type of data-enabled object to be deployed in the wild, as one of the members said: ``If I were [to] be in the same situation, I think I would be okay with this camera too'' (G5). However, some team members had a more negative attitude toward the camera: ``
I still would not feel comfortable having it deployed in my house, especially after seeing how much data is gathered'' (G18). No matter how the design teams responded personally toward the use of the camera, the CPT enabled a privacy-sensitive situation that allowed the design teams and their participants to experience, investigate, discuss, and reflect on privacy preferences and related design requirements.

\subsection{From Reflection to Design Action}
Some members of the design teams not only revised their own understanding of privacy after reflecting on the different reactions of their participants, they even proposed their own privacy design requirements: ``Thinking about my own privacy, I would not mind having to disclose some private data provided it is for a greater cause. I would prefer to know what the outcome of the data is and what [it might] be used for before allowing it to be disclosed'' (G18). Some team members also reflected on the differences between theory and practice when examining the actual reactions of their participants. For example, they saw that some participants were resistant to being exposed to any form of data collection (G8), whereas others appreciated it (G2). Some teams also experienced how fast their participants adapted to the presence of data-enabled objects: ``I think participants got used to the camera very fast...people are often unaware of what data is collected of them and [what]can be done with this data. I think one of the biggest problems of IoT is that people bit by bit surrender their privacy'' (G9). Another member commented: ``Privacy was ethically right, but in practice, participants responded differently than expected'' (G11). The following sections present further reflections by the design teams on their changing perspectives and on proposing design actions.

\subsubsection{Changing Privacy Design Perspectives.} 
The design teams found that every new tactic they might come up with to protect privacy could lead to yet another privacy problem (G17). Privacy design, they found, can be broader than they had previously imagined (G6) and even more complex. For example, as the groups added additional sensors to expand the environmental awareness for building a non-intrusiveness system, they often found that these sensors still recorded participants' actions without permission. One of the home participants in the deployment of G3 shared that \textcolor{minor}{``...the camera still recorded my feet when I was walking on the stairs. The distance sensor did only sense whether there was someone in front of the shoe rack. This gave me a somewhat unsafe feeling.''} \textcolor{revision}{A G14 member also reflected, ``[We] had never actually realized how dangerous sensors in a home can be and how much data can be gathered from just observing someone's patterns.''}
The team members in G1 realized that privacy cannot be well-designed or well-defined, and that even just the simplest data can raise crucial privacy concerns: ``Especially in a home, privacy is an aspect that should get a lot of careful attention. This holds not only for the actual taking [of] pictures but also for all the other implicit information that is gathered like living patterns and the activities of housemates that the other housemates might not know of'' (G14). 
The design teams described how they changed their perspectives from solving privacy into exploring privacy: ``[Before] I thought of privacy as a somewhat objective idea. Something either breaks privacy concerns or it doesn't. However, here I learned that multiple reactions to privacy are possible, and this is something that needs to be taken into account when designing data gathering systems'' (G18). In the beginning, the teams struggled with searching for an exact solution to dealing with privacy. Then, as one member discovered, there did ``[not necessarily] have to be a single solution'' (G9). Privacy can be and needs to be explored; it is not something easily solved in one go.

\subsubsection{Proposing Privacy Design Principles.}
The teams were able to propose their own requirements for privacy design, and identified that the most important principle is \textit{participants' empowerment}. As one member from the team G13 put it: ``I believe the key is to involve the user as much as possible.'' While privacy issues could never be fully resolved no matter what the design teams built, they often found that the important thing was to involve their participants in the research process from beginning to end. As one team member (G5) mentioned, ``people were more willing to have a picture of them taken when they feel that they can refuse / disable the system at any time.'' By empowering their participants with capabilities to design and control the data-enabled objects, the teams were able ``to figure out their concerns and take them into account when designing the system'' (G1). Finally, every design team was capable of proposing their own revision plan for a future data-related project, as expressed by a member of G11: ``I think that in future work I will definitely do a pilot test with participants first, before going to collect actual data. This hopefully will lead to better data gathering on the design case instead of the observations on privacy only.'' 

\subsection{Summary}
The CPT successfully facilitated 18 different home studies with data-enabled objects, and it allowed the design teams to carefully investigate privacy-specific requirements and design impacts with participants in the wild. Although some of the design teams mentioned that \textcolor{minor}{the} CPT sometimes had unexpected hiccups during the data collection (G1, G6), they were still generally positive about the provided toolkit and the chosen approach. Through actual implementation, the process allowed them to ``open their eyes'' (G11) to see how privacy is intertwined with design research practice (G5) and how they could finally synthesize privacy-aware designs (G15). Ultimately, their reflections on their studies allowed us to interpret their understanding of privacy design and how it evolved throughout the design research activity.


\section{Discussion}
In this section, we revisit our overall study to respond our four research questions. First and second, we discuss how we support\textcolor{minor}{ed} design researchers by using a toolkit for designing and using data-enabled objects, and the role \textcolor{minor}{that} our toolkit \textcolor{minor}{played} (RQ2). Third and fo\textcolor{minor}{u}rth, we revisit how the 18 student design \textcolor{minor}{teams} use\textcolor{minor}{d} our toolkit to explore the privacy design space for data-enabled objects and how the five design aspects can \textcolor{minor}{suggest future design directions} for addressing the privacy challenges \textcolor{minor}{of} data-enabled objects in design ethnography (RQ1 and RQ3). Fifth, we reflect on the overall study to discuss our changing perspective on privacy (RQ4). Finally, we discuss the limitations of our work and recommend future steps in design research working with data-enabled objects. 

\subsection{A Toolkit for Bottom-Up Privacy Design Space Explorations}
\textcolor{revision}{Ledo et al.~\cite{Ledo_2018_CHI} identified and proposed a lens for evaluating toolkits used in HCI research. Based on their lens, we examined how our toolkit could be used as an approach for supporting researchers in identifying broader research themes (i.e., privacy design space, in this paper).} \textcolor{minor}{We identified that the CPT can be used for a type of ``design space exploration~\cite{Ledo_2018_CHI}.''} \textcolor{minor}{In this category,} examples by Houben et al.~\cite{Houben_2015_CHI} and Marquardt et al.~\cite{Marquardt_2011_UIST} represent top-down explorations in which the researchers created a collection of design spaces from literature reviews, then demonstrated the new applications that are enabled by the respective toolkit. In contrast, our proposed CPT enables a bottom-up exploration similar to TaskCam~\cite{Boucher_2018_CHI}. For example, the CPT provides open design templates that can be used to explore expression, data control, and interaction of data capture with regard to legibility, agency, and negotiability~\cite{lindley_2020_CHI, mortier_2014, victorelli_2019} (see Section 5.2). These templates allow design teams to begin with a similar type of data-enabled object (a camera), explore additional data-enabled capabilities by making external products, and identify their own design strategies. As such, they built similar but different cases that together construct a collective example of how to enrich the design space for data-enabled objects. All in all, the use of bottom-up explorations to generate a collection of examples ultimately results in potential design elements that can be directed at the design space of privacy-aware data-enabled objects.

\subsection{Controllable Interventions to Embody \& Investigate Privacy in the Wild}
The CPT, designed with local connectivity, interaction API, and \textcolor{minor}{D}ata \textcolor{minor}{C}anvas, provided a fundamental data infrastructure that supported the teams in designing and using privacy-sensitive, camera-based, data-enabled objects in any context under a controllable intervention. As privacy can be very subjective and context-specific, the CPT enabled a designerly intervention by facilitating a lived, yet ethnographically observable, experience in a domestic context (see Section 5.3). However, the crucial point is that such an intervention must be controllable by the residents in the home in terms of its privacy impact. This means that, during the explorations, all home participants could reject data collection and redact personal data at any point in the studies---and could do so based on informed privacy choices. The data collection process that is part of the CPT supported our design teams in building such a controllable intervention. 
Based on the toolkit, the teams tested their data-enabled objects in a real-life domestic context. They came to understand the potential and real privacy impact of their design choices directly from in-situ deployments. Their reflections show that, for example, they became aware of differences between theory and practice, realized that multiple reactions to privacy are possible in design, and were able to propose their own takes on privacy requirements and design insights.

These insights and reflections would not have happened without exploring privacy in real-life through a designerly practice. Privacy issues can result from the interplay of data collection interests, data ownership, and the protection of sensitive information. This interplay has developed historically as a slowly-emerging issue with widespread implications. As more and more data are collected and stored, connections and relations emerge that are complex and difficult to predict for (the often involuntary) data-sharing individuals. For example, many cases can be seen in the context of recent data breaches by Amazon Echo~\cite{amazon_2019} and Google Home~\cite{google_2019}. \textcolor{revision}{To carefully deal with the privacy issues that occur during data collection, prior research has emphasised the need for addressing all kinds of details even when just building a simple function such as notification (i.e., privacy notice~\cite{Schaub_2015_SOUPS} and privacy choice~\cite{Feng_2021_CHI}). 
Researchers have also discussed the need for gaining consent from participants during the HCI research process, thus ensuring what they call ``ethical HCI research~\cite{Brown_2016_CHI}.'' Their work has emphasised on empowering participants by allowing them control over privacy-related situations and the chance of reconsiering their privacy-related decisions~\citep{Luger_2013_Ubicomp}. These lines of work all highlight an important design considerations, that is to empower participants' agency during the data collection.}

Even if co-design with participants can help researchers understand privacy requirements before they implement or deploy data-enabled objects, this understanding will always be approximate at best. Many aspects and requirements for privacy can only be revealed by in-situ deployments and reflective practice. Exploring privacy in practice enabled the design teams to become (more) aware of different subjective realities in a given home. As a result, every student in the course was able to start discussing and approaching privacy-sensitive contexts in a more practical and, perhaps, bolder way. As more connected or smart products enter our everyday lives, addressing privacy concerns through reflective and multidimensional negotiation becomes increasingly important~\cite{mortier_2014}. The prevalent notion of whether to build these connected products more restrictively in the future is not necessarily a given. Instead, such a decision should be made based on reflection in action~\cite{schon_1983}. Many of the design team members found after the exploration that they appreciated this chance to actually work in practice, as reflected in one member's comment: ``I now have the confidence and knowledge for dealing with (privacy) sensitive data'' (G11).


\subsection{\textcolor{revision}{Giving Contextual Awareness to Data-Enabled Objects}}
Our toolkit enables designers to build a system of data-enabled objects in which \textcolor{minor}{the} multiple objects can share data, extend perceptions, and \textcolor{revision}{flexibly perform their data-capturing tasks concerning different situations in a home}—all with the purpose of extending the original Peekaboo Camera~\cite{Cheng_2019_DIS} with additional contextual sensing in order to know when it is appropriate to take photos and to interact with participants. 

\textcolor{revision}{This can be done by giving data-enabled objects the contextual awareness of knowing ways of capturing data in the right contexts, in the right amount and at the right time. For example, with the support of the toolkit, the design teams built and deployed various sensors into different locations to capture contextual data in a home and at the same time to control the camera to capture context-relevant photos (e.g., capturing photos of clothing and shoes patterns taking place when participants were entering the home entrance and not standing in front of the camera (G18)). 
Not only the design teams but also the home participants could enrich this contextual awareness of data-enabled system. For example,} 
some of the objects were built as an interactive artifact, designed to interact with the participants and trigger a function (e.g., a photo-taking trigger key holder in G18); and some of the objects were designed as an emergent data control interface, employed to remind the participants how they could intervene in the system at any time (a red button in G5). \textcolor{revision}{Participants became part of these data-enabled systems by intervening in the data collection through allowing more or fewer photos to be captured.} 

Inevitably, of course, the pre-defined contextual meaning is interpreted and cannot always be true. For example, one of the objects accidentally detected the motion of a cat, but it was recognized by the system as a human walking by. 
\textcolor{minor}{Design teams} cannot perfectly interpret every situation that occurs. \textcolor{minor}{Nevertheless,} we still found that some of the participants were amazed by the systems designed by our design teams (G18) and asserted that their design represented an ``intelligent system''. Although each data-enabled object was simply integrated with a sensor (e.g., a motion detector, vibration or heat sensors), using no advanced technology (e.g., artificial intelligence), people saw them not individually but as a data-enabled system that ``magically'' supported the camera to take photos and interact with them in an appropriate manner (G18).

\subsection{Giving Product Qualities to Data-Enabled Objects}
Prior work has proposed the important steps for researchers to better sustain and maintain the data collection process~\cite{tolmie_2010, lovei_2020_DIS}; however, limited studies have further investigated the product qualities that data-enabled objects and systems need in order to interact with the research participants. The five design aspects (form, observational perspectives, notification \textcolor{minor}{of data capturing}, \textcolor{minor}{interaction mode of} data \textcolor{minor}{capturing}, and data processing) suggest various design qualities that future researchers can refer to in designing the interaction between their data-enabled objects and systems and the research participants. For example, \textit{Form} can be designed using different shapes that can convey different messages that trigger social interaction among participants, thus influencing the data collection. \textit{Observational perspective} can also be seen as a quality of designing sensing sights for different types of data-enabled objects. While designing an appropriate observational perspective can directly reduce participants' anxiety about being observed, at the same time, it also allows design researchers to explore alternative and creative perspectives for observing the field, such as the object-only perspective. \textit{Interaction \textcolor{minor}{Mode} of data capturing} and \textit{Notification of data capturing} can be considered as a means of designing for agency and negotiability, allowing data-enabled objects to be better aware of different situations in the field and to be able to react flexibly; furthermore, additional data anchors are needed to add meaning to the interaction. Finally, \textit{data processing and mapping} also support negotiability and enhance readability of \textit{boring data}, allowing a greater degree of sense-making and more accurate mapping of semantics to the data; this can lead to further interpretation and re-filtering of the data. In response to the potential privacy challenges identified in the beginning of the paper (engagement, empowerment and enactment), we found that choices made with regard to these five design aspects can provide different strategies for designing data-enabled objects so that they can be better integrated \textcolor{minor}{into} people's daily activities, empower\textcolor{minor}{ing} both \textcolor{minor}{the} people and \textcolor{minor}{the} data-enabled objects with a shared agency, and \textcolor{minor}{thus leading to} better enact\textcolor{minor}{ment of the} design ethnographic task. 

\subsection{Designing Privacy as a Driver }
\textcolor{revision}{The concept of privacy can be traced back centuries. One of the famous early moments in this history was the 1604 English legal decision known as Semayne's case, which established important concepts about a person's right to deny entry to privately owned property. Another occurred around the 1760s in colonial America, with the strong reactions against search warrants, imposed by the British authorities, that played a role in inspiring the American Revolution. Another was a seminal essay titled The Right to Privacy, published in the Harvard Law Review in 1890, that established the well-known statement, ``privacy is the right to be let alone''~\cite{warren_1890_law}. 
This argument laid the conceptual foundation for researchers in seeing privacy as shaping a limited access of self, 
 or empowering an individual control over personal information\textcolor{minor}{, within a range} between limited access and fully access~\cite{Nippert_2007_ijd, westin_1967_book}.} At the same time, \textcolor{minor}{the} researcher Solove \textcolor{minor}{did} a review \textcolor{minor}{of} conceptual privacy from \textcolor{minor}{a} legal \textcolor{minor}{perspective} and interpreted that ``privacy cannot be consolidated into any single conception; [instead] they cluster together certain of the conceptions''~\cite{solove_2002}. Shifting from the law aspect to design, in 2009, Cavoukian proposed the \textit{Privacy by Design} framework (data protection through technology design) to highlight that any data-related design should make privacy \textcolor{minor}{a} default requirement; rather than a trade\textcolor{minor}{-}off. This means \textit{Privacy by Design} should allow users  to have full \textcolor{minor}{access to all functions,} loosing \textcolor{minor}{no benefits, even if they refuse to allow the company to gather any data on them}~\cite{cavokian_2012, cavoukian_2010}. Their framework invoked a user-centric perspective and \textcolor{minor}{called for} designing privacy through ``Proactive not Reactive; Preventative not Remedial'' design. Such \textcolor{minor}{a} framework was adopted by the International Assembly of Privacy Commissioners and Data Protection Authorities in 2010, and the European GDPR regulation in 2016. The intention behind the framework is to present a perspective of privacy design as a data prevention design. 

\begin{figure}[h]
\centering
  \includegraphics[width=1\columnwidth]{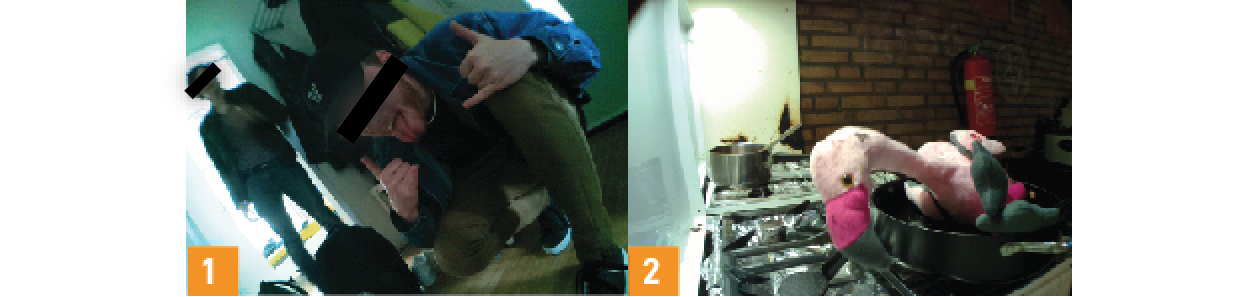}
  \caption{In some design cases from the teams, their participants used data-enabled objects for different purposes. 1) G18 found their participants use\textcolor{minor}{d} data-enabled objects to \textcolor{minor}{take} selfie\textcolor{minor}{s} when \textcolor{minor}{a} friend \textcolor{minor}{was} visiting. 2) G3 found that their participants put funny toys in front of the data-enabled objects to ``trick'' \textcolor{minor}{them}.}~\label{fig:unexpected}
\end{figure}

In these discussions about privacy, privacy design seems to be often interpreted as a design that ultimately constrains the experience of users (or participants). In this mindset, privacy becomes an obstacle for developers, designers or researchers who need data, and they see users (or participants) in an opposite position that tends to limit data collection. However, in our explorations, several student cases \textcolor{minor}{showed} that participants are not always in an opposite position and can become a collaborator or even a data curator\textcolor{minor}{, capturing} and shar\textcolor{minor}{ing} unexpected photos to the data-enabled objects. For example, teams G18 and G3 found that their participants used the data-enabled objects to actively capture more photos than they expected. \textcolor{minor}{Team} G18 found that their home participants made a group selfie when a friend visited (see Figure~\ref{fig:unexpected}-1). \textcolor{minor}{Team} G3 received a series of toy-in-a-pot photos captured by the participants\textcolor{minor}{,} who \textcolor{minor}{wanted to} intentionally trick the data-enabled objects (see Figure~\ref{fig:unexpected}-2). Furthermore, any intervention that participants made \textcolor{minor}{was} not preventing data collection; instead, it \textcolor{minor}{was} enriching data collection with additional semantics. For example, \textcolor{minor}{T}eams G5 captured additional information about \textcolor{minor}{an} unavailable photo-taking moment when the participants \textcolor{minor}{were} having intimate interactions with others (see section 6.5.4). In these cases, the participants became co-ethnographers, and as such were themselves curious about what perspective\textcolor{minor}{s} could be captured, and intentionally intervened and organised \textcolor{minor}{a} scene for the data-enabled objects to capture. 
As such, the student design teams were able to receive more interesting data to discuss with the participants. 

All of these examples happened because the student design teams not only took privacy into account in an earlier stage but also invited their home participants to be involved in the design process. Through the CPT-scaffolded design \textcolor{minor}{(i.e., built-in connectivity, programmability and data-sharing interfaces)}, \textcolor{minor}{the CPT provided the researchers, participants and data-enabled objects the flexibility and the control to adjust the data flow}---before, during, and after the data collection. The CPT-scaffolded design not only empowers participants to always be the guardians of their own privacy \textcolor{revision}{but also invites different stakeholders to co-design the system with other researchers. For example, P47 in G9 highlighted that ``everybody receiving a Peekaboo module to build a system around is a very good opportunity to make a somewhat bigger system.'' By looking at how other different groups redesigned their Peekaboo Camera, the groups could also learn from each other and adapt their own designs.} 
At the same time, the CPT-scaffolded design empowered the design researchers to \textcolor{revision}{become more sensitive to privacy design and to} explore different \textcolor{minor}{design} opportunities \textcolor{minor}{with their participants.} \textcolor{revision}{For example, when G4 deployed a pressure sensor inside a couch for detecting how long participants sat on the couch, G4 specifically decided that ``the weight of the person should not be kept'' and that only the data consented to (i.e., the time period) could be captured. With our approach, the design teams were able to work together with the participants to carefully determine the appropriate balance of privacy requirements (e.g., what data to keep and what data not to keep).}

\begin{figure}[h]
\centering
  \includegraphics[width=1\columnwidth]{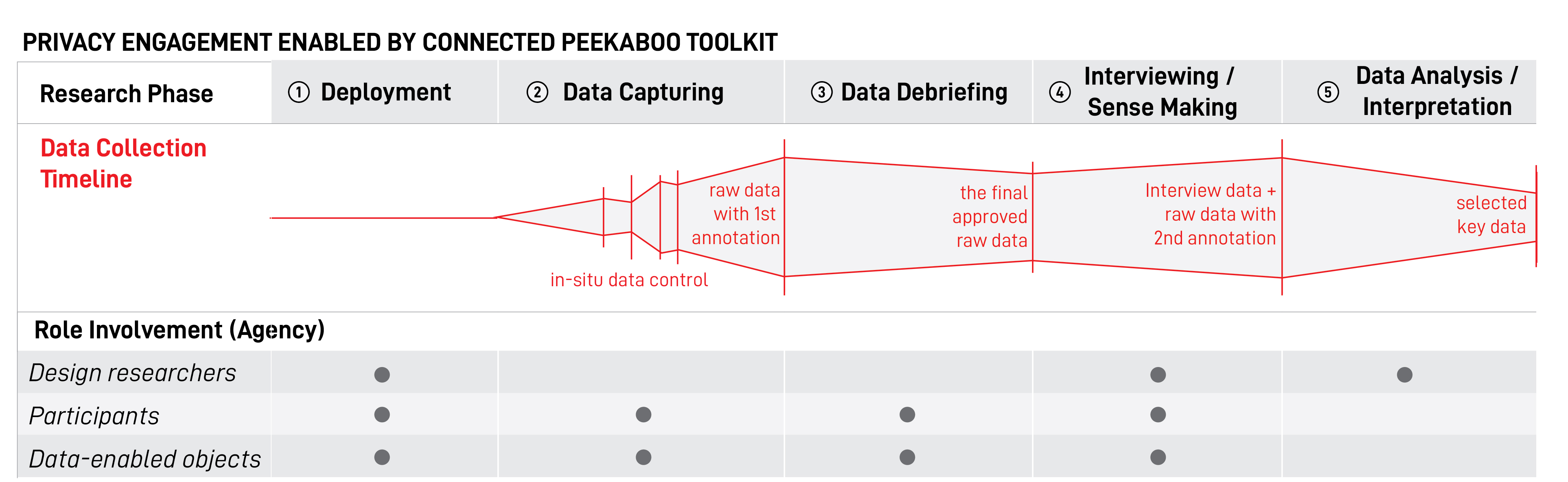}
  \caption{
\textcolor{minor}{Privacy Engagement} enabled by the Connected Peekaboo Toolkit (CPT)\textcolor{minor}{. This figure} shows how data collection \textcolor{minor}{was} controlled by different \textcolor{minor}{agents} (design researchers, participants and data-enabled objects) \textcolor{minor}{through} the five research phases (deployment, data capturing, data debrief\textcolor{minor}{ing}, interview\textcolor{minor}{ing} and data analysis). 
}~\label{fig:intervention}

\end{figure}

Compared with related ethnographic approaches such as thing ethnography~\cite{Giaccardi_2016_DIS} and data-enabled design~\cite{Bogers_2018_DIS, Bogers_2016_DIS}, in which the participants usually play a passive role, being observed by the researchers and the data-enabled objects\textcolor{minor}{, the} CPT-scaffolded design exploration \textcolor{revision}{facilitates privacy engagement for every actor participating in the study. Researchers and home participants all} take on an active role in deciding \textcolor{minor}{what} they wanted to \textcolor{minor}{observe} during the deployment, data capturing, data debrief and interview. \textcolor{minor}{(Privacy engagement according to different roles is detailed in Fig~\ref{fig:intervention})}. 

As a result, the participants can work collaboratively and begin to share their \textcolor{minor}{lives} with the data-enabled objects and even to play with them in order to explore what different kinds of perspectives the objects can capture. The data collection enabled by CPT can be selected and enriched with meaningful annotations. 

\textcolor{revision}{Our work explicitly empowers participants with ``legibility, agency and negotiation (HDI)~\cite{mortier_2014}'' before, during and after the data collection. However, we are not claiming that the CPT allows all design teams to design perfectly for legibility, agency and negotiation during the data collection. Instead, our exploration with design teams has shown that when a design team comes up with a new design tactic to address privacy, such design tactic can create the opportunity to uncover new privacy challenges (see section 7.3.1). Nevertheless, these privacy requirements may not be easily identified without encouraging people to actually experience the data collection (see section 7.2). Our work encourages more future design research to continue shaping controllable intervention (see section 8.1) for exploring wider and deeper in-situ privacy requirements of participants.} The results gained from CPT-scaffolded explorations change\textcolor{minor}{s} our perspective to \textcolor{minor}{one that} frame\textcolor{minor}{s} privacy less as an obstacle and more as a driver. We assert that privacy design should not be about designing an interaction for blocking data; it should be about designing privacy as a driver to empower participants, researchers and data-enabled objects to have a shared agency for shaping the data collection better. The results and reflection\textcolor{minor}{s} from the student teams support this from an empirical standpoint. 


\subsection{Limitations and Future Work}
We are aware of limitations in this study. 

First, all of our teams' design explorations were situated in a home context, which, by definition, is a personal, private, and restricted space for experimentation. The users \textcolor{minor}{were}, however, in full control of this space; they \textcolor{minor}{could} decide about their privacy needs and act accordingly. Different issues might arise when deploying data-enabled objects into other contexts such as an office space or a school. Both are semi-public spaces that are governed by additional constraints, such as work regulations and special ethical provisions for vulnerable or dependent populations. Not only could privacy-sensitive data be recorded differently, but also additional considerations (such as collective consent, the different needs of various occupations, and different kinds of sensitive content) \textcolor{minor}{would} need to be addressed. 

Second, while the current design spaces related mainly to the design of camera-based data-enabled objects, different perspectives could be found by using different types of sensor-based data-enabled objects. For example, new modalities such as sound might lead to discussions about whether a data-enabled object should capture complete human sentences or random noise. Additionally, future research should explore more combinations and choices based on our current five design spaces for data-enabled objects when deploying them into more varied contexts. However, that does not mean that we see a definite and ideal set of combinations for designing a data-enabled system. Instead, we see the five design spaces suggesting a framework for designing diverse data-enabled objects with different forms, features, functions, and data-processing capabilities. 

Third, another limitation is that our invited design teams, all second-year undergraduate students, were relatively inexperienced in research methodology. Guiding the students took considerable effort, because they had to both develop an understanding of the challenges and perform design research activities simultaneously. For example, we provided extensive briefings to the design teams prior to all tasks. The process was also structured in such a way that they could test their prototypes of data-enabled objects in a safe environment. In hindsight, we were impressed with the design teams' professional attitudes and careful steps towards making privacy-conscious design choices. We observed that the students made bigger and bolder steps as their confidence grew throughout the project, without losing their initial creativity.

\textcolor{revision}{Fourth, although the CPT was introduced in a design course with 75 bachelor-level students, it was used in more contexts than in the classroom. In fact, the CPT, with its three fundamental design features (built-in connectivity, programmability and data sharing interface), allowed the design teams to flexibly build their systems of data-enabled objects in different contexts. We have seen that the design teams prototyped their systems at home, tested their systems at school, and deployed their systems in a real-life context. The goal of the CPT is to bring privacy awareness to design researchers (i.e., students, in our study) by engaging them early-on in the design process and in shaping the data flows with respect to privacy. We believe the CPT can provide future researchers with great flexibility in exploring appropriate interactive systems of privacy-aware data-enabled objects for studying different contexts (e.g., schools, hospitals, train stations). }


\textcolor{revision}{Fifth, we found that the 18 written reports and the 75 individual reflections captured both their personal subjective experiences and also reflections from the in-class feedback and discussions with others. 
For example, P49 in G15, reflecting on feedback from their tutors in the course, noted that their system design could be used for different purposes: ``We got feedback that our product could also be used to analyze eating habits of pets. Which was a very nice idea in my opinion, and not something I would have thought of at first.'' However, we are aware that the written reports can only present design teams' after-thoughts, reflections and considerations. Future studies might consider conducting in-class observations and interviews to study the students' unaware behaviours, thoughts and in-situ reflections for privacy design.}

Finally, our eight-week design explorations\textcolor{minor}{,} from designing to prototyping to deploying, still resulted in severe time pressure for the design teams. Due to the limited time, we noticed that some of the design teams often built similar forms and functions for data-enabled objects to investigate the same context. For example, some design teams limited themselves to building bottle-based or stove-based sensors for studying kitchen activities, not considering that many other everyday objects could also be used. Therefore, a future work could provide some templates or design explorations that focus on inspiring designers to identify diverse forms and functions, and speculate about contextually-relevant data before designing their data-enabled objects.

\section{Conclusion}
\textcolor{revision}{Data has become a new material for design researchers to use in understanding people's daily lives from different perspectives~\cite{Bogers_2016_DIS, Bogers_2018_DIS}. However, with more and more sensor-augmented research objects or ``data-enabled objects'' being employed in the design ethnographic process, privacy issues are on the rise. To deal with these privacy issues, we developed the Connected Peekaboo Toolkit (CPT), a combination of open-source hardware and software, for supporting design researchers in building privacy-aware camera-based data-enabled objects for different ethnographic contexts. With the CPT, we are now in a position not only to adapt the data collection to the research needs, but also to undertake a careful assessment of the privacy balance, and to do so with the direct involvement of participants. Our results of applying the CPT in Design Research show that} \textcolor{minor}{it} successfully support\textcolor{minor}{ed} 18 different teams of student designers (75 students in total) to build 18 design explorations\textcolor{minor}{, which were} deployed and evaluated in different households. By using the CPT to design specific data-enabled objects, \textcolor{minor}{the} design teams \textcolor{minor}{could} explore privacy design for data-enabled objects through reflective practice. We compared their design outcomes and reflections to come up with a synthesis of various design aspects that can be considered in seeking to balance the tension between privacy and ethnographic needs. 

\textcolor{revision}{In summary, this work offers three major contributions.} 

\textcolor{revision}{First, we have conceptually introduced ``data-enabled objects'' as a new type of research object that extends the quality of research products~\cite{Odom_2016_CHI} and captures additional contextual data for design ethnography. We also have identified general data privacy challenges during the design ethnographic process. Three data privacy challenges---\textit{engagement, empowerment,} and \textit{enactment}---guide design researchers in addressing data privacy issues by employing participant-in-the-loop data collection.}

 
\textcolor{revision}{Second, we present the Connected Peekaboo Toolkit (CPT), that provides design researchers three privacy-aware design features---\textit{built-in connectivity, programmability,} and \textit{data sharing interfaces}---for practically designing an in-situ data-capturing collaboration between researchers, data-enabled objects and participants for design ethnography. The CPT allows design researchers to customize the data-capturing system to meet the privacy and research needs of both participants and researchers. Our results show that many privacy issues can only be found through in-situ deployment. The CPT facilitates a reflective process for design researchers to contextualize data privacy needs for different participants and homes, and to effectively adapt data collection to meet these needs.} 


\textcolor{revision}{Third, we describe a framework of five design-space elements---\textit{forms, observational perspectives, notification of data capturing, interaction mode of data capturing,} and \textit{data processing}---for design researchers to use in designing and deploying data-enabled objects in capturing and using data for design ethnography with the respect to privacy. The framework elements were identified from the 18 design explorations, and they provide future design researchers with actionable guidelines to design and facilitate participants' engagement with data collection before, during and after design ethnography. } 

\textcolor{revision}{Taken together, our contributions present an optimistic outlook on privacy-aware practices in design ethnography. We claim that privacy need not be an obstacle but can be a driver for design researchers to (1) identify more varied privacy needs, opportunities, and considerations for designing privacy-aware ethnographic research data, and (2) engage participants in data-enabled, deliberately participatory ethnographic practices.}

\bibliographystyle{ACM-Reference-Format}
\bibliography{main.bib}

\appendix


\section*{Supplementary material}

APPENDIX

Appendix 1. Overview of the 18 Design Projects: Designing a System of Privacy-Aware Data-Enabled Objects for Home Ethnography

| Groups | Group Members | | Project Name | Target Context | Total Days | Name of the Data-Enabled Objects | Additional Designed Gadget | Sensors | Deployed Location | Numbers of Data Items | | Function | Purpose |
|---|---|---|---|---|---|---|---|---|---|---|---|---|---|
| | Numbers | Background | | | | | | | | Total Photos | Total Data Points* | | |
| G1 | P16 | Industrial design | Cycle | Shared Balcony | 3 | Peekaboo Cam | - | - | Balcony | 26 | 493 | Take photos a few seconds after the presence of a person is detected. | Capture ethnographic photos for studying activities before people get on a bike. |
| | P22 | Industrial design | | | | | Human detector | PIR sensors | | | | Detect human presence | To capture photos when someone is in front of the Cam. |
| | P34 | Industrial design | | | | Sensor Box | - | Ultrasonic sensors | Toilet | | | Detect human presence | To know when people use the toilet. |
| | P39 | Industrial design | | | | Sensor Box | - | Infrared sensors | Backdoor | | | Detect human presence | To know when people leave by the backdoor |
| G2 | P66 | Mechanical engineering | Improving Student Home Hygiene | Shared Kitchen | 7 | Peekaboo Cam | - | - | Kitchen Cabinet | 19 | 246455 | (1) Take photos when detecting plates on the Kitchen Flat for more than 25 minutes. (2) Delete photos when the Kitchen Flat detects no dishes and human presence. (3) Save photos when the Kitchen Flat detects dishes and no human presence. | Capture photos of kitchen culprit. |
| | P53 | Psychology Technology | | | | RFID reader | - | RFID sensor | Kitchen Door | | | Scan the ID tag to delete the last photo taken from the Peekaboo Cam | To prevent taking photos of people who are not the culprit. |
| | P54 | Psychology Technology | | | | Sensing Kitchen Flat | - | Force sensor | Kitchen Counter | | | Detect dishes on top of the flat | To know when the dishes are unclean and placed on the kitchen counter. |
| | P55 | Psychology Technology | | | | | | | | | | | |
| | P45 | Software Sciences | | | | Sensing Kitchen Wall | - | Ultrasonic sensor | | | | Detect human presence | To know when people use the kitchen |
| | P58 | Psychology technology | | | | | | | | | | | |
| G3 | P13 | Industrial design | Three Pans and Pots | Shared Kitchen | 4 | Peekaboo Cam | - | - | Kitchen | 85 | 11672 | Take photos when someone is walking away from the stove. Notifying people when they are walking in front of the camera. | People will not be in the photo, because of identified privacy needs, and they will also be warned. |
| | P17 | Industrial design | | | | Sensing Stove | - | Distance sensors | | | | Detect human presence when they are in front of the stove. | (1) To know when participants using the stove. (2) To control the camera to take picture. |
| | P51 | Psychology technology | | | | Sensing Door | - | Distance sensors | | | | Measuring the number of people walking in and going out of the door | To know how many people are in the room at any one time. |
| | | | | | | Disabled Plate | - | Infrared sensors | | | | Disabling photo-taking when someone puts their hands in front of it. | Allow participants control their own privacy in a certain way. |
| G4 | P6 | Industrial design | Taboa | Shared Dining Room | 4 | Peekaboo Cam | - | - | Dining Room | 67 | 6077 | Take photos at the moment when all of three other sensing objects are activated: (1) detecting someone on the chair, (2) detecting the audio wave at high peak, (3) detecting enough lights in the room | Capture ethnographic photos for studying human activities and conversations at the dining table |
| | P9 | Industrial design | | | | | Privacy Filter | - | | | | Filter unauthorised areas to be captured | Make sure the Camera captures photos in an appropriate frame |
| | | | | | | Sensing Cushion | - | Load cell amplifier | Dining Chair | | | Detect human sitting on the chair (true/ false) | Make sure the Camera capture photos of people; however, ''The weight of the person should/would not be captured''. |
| | P21 | Industrial design | | | | Sensing Table | - | microphone | Dining Table | | | Detect the audio wave changes | Make sure the Camera captures the photos during people's conversations |
| | | | | | | Sensing Table | - | Lighting sensors | | | | Detect the lights in the room | Make sure the Camera is enough light for the Camera to take photos. |
| G5 | P12 | Industrial design | Knock Knock Who's There | Shared Home Entrance | 3 | Peekaboo Cam | - | - | Home Entrance | 27 | 493 | Take photos when participants are walking out of the house. The cam would delay photo taking for 10 seconds. | Capture ethnographic photos for studying human activities when going out of the home. At the same time, the camera delays the photo-taking to give participants enough time to prepare beforehand. |
| | P20 | Industrial design | | | | Red Button Box | - | - | | | | Delay the Cam to take photos for 10 seconds. | To give participants power to control the camera |
| | P50 | Software sciences | | | | Sensing Mug | - | Pressure Sensors | | | | Detect humans walking into the house or walking out of the house. | To know when people come into or go out of their home. |
| | P61 | Psychology technology | | | | | | | | | | | |
| G6 | P1 | Industrial design | Privashoe | Shared Shoes Rack | 7 | Peekaboo Cam | - | - | Home Entrance | 319 | 6347 | Take photos of the shoes rack when participants are not present. | Capture ethnographic photos for studying changing shoe patterns on the shoes rack everyday. |
| | P2 | Industrial design | | | | | Human detector | Distance sensors | | | | Detect human presence | To prevent photo taking when someone is in front of the Cam |
| | P36 | Industrial design | | | | Switch Button | - | - | | | | Disable photo taking for a period | To give participants power to control the camera |
| | P37 | Industrial design | | | | Sensing Door Bell | - | Sound Sensors | Door Bell | | | Detect the moment when the doorbell rings | To know when people are coming in and opening the door |
| G7 | P24 | Industrial design | Mimic | Shared Dining Room | 15 | Peekaboo Cam | - | - | Dining Room | 58 | 12505 | Take photos of the dining table when participants stand up from a chair and leave the table. | Capture ethnographic photos for studying the object arrangements when participants leave the dining table. |
| | P25 | Industrial design | | | | | Flag | - | | | | Raise a flag when the participants entering the area. | To notify participants that the system is detecting their presence. |
| | P69 | Applied mathematics | | | | Sensing Table | - | Ultrasonic sensors | Dining Table | | | (1) Detect humans entering and existing the area around the dining table. (2) Measuring the number of times participants enter and leave the area. | (1) To prevent participants being captured by the Cam. (2) To know how many people are in the room at a particular time. |
| | P70 | Applied mathematics | | | | Sensing Chairs | - | Infrared sensors | Dining Chair | | | (1) Measure how long do participants sit on a chair (2) Detect when participants stand up from a chair | To know how long do participants sit on a chair To command the Cam to capture photos when someone sits down and stands up from a chair. |
| | P71 | Computer science | | | | Switch-off button | - | - | Dining Table | | | Disable the photo-taking immediately. | To give participants power to control the camera. |
| | P61 | Electrical engineering | | | | | | | | | | Capture ethnographic photos when the participants are sitting on the couch and the TV is turned on. | Capture ethnographic photos of object arrangements on the TV table for studying activities occurring when people are sitting on the couch and watching TV. |

| Groups | Group Members | | Project Name | Target Context | Total Days | Name of the Data-Enabled Objects | Additional Designed Gadget | Sensors | Deployed Location | Numbers of Data Items | | Function | Purpose |
|---|---|---|---|---|---|---|---|---|---|---|---|---|---|
| | Numbers | Background | | | | | | | | Total Photos | Total Data Points* | | |
| G8 | P62 | Electrical engineering | The Couchpota toes | Shared TV table | 2 | Peekaboo Cam | - | Distance sensors | Living Room | 14 | 12086 | Detect human presence | To prevent photo taking when people are in front of the Cam. "This is only to avoid capturing people on camera, so we did not save this data. We placed the distance sensor on top of the peekaboo such that it did not get triggered by anything on the table (at least not by anything smaller than a water bottle)." |
| | P64 | Electrical engineering | | | | Sensing TV | - | LDR sensors | TV | | | Detect when the TV is on and off | To trigger the camera; however, "It doesn't send any data when the tv is off." |
| | P67 | Mechanical engineering | | | | | | | | | | | |
| | P72 | Computer science | | | | Sensing Pillow | - | Pressure Sensors | TV Couch | | | Detect how long participants sit on a couch | To know how long of people sitting on |
| G9 | P4 | Industrial design | Who's Home | Shared Home Entrance | 7 | Peekaboo Cam | - | - | Home Entrance | 71 | 1178 | Capture photos when the door opens | Capture ethnographic photos for studying activities when people are entering and leaving the home. |
| | P18 | Industrial design | | | | | Lens Filters with three levels (no-filter; a light filter; heavier filter) | - | | | | Filter unauthorised areas to be captured | To prevent taking photos of the area which are not allowed to be captured. |
| | P19 | Industrial design | | | | Sensing Door Mats | - | Pressure Sensors | | | | Detect humans walking into the house or walking out of the house. | To know when people come into or go out of their home. |
| | P31 | Industrial design | | | | Sensing Door | - | accelerome ter | | | | Detect door movement | To know when the door is opened. |
| | P47 | Software sciences | | | | Button | - | - | | | | Delete the captured photos immediately | To give participants power to control the camera |
| G10 | P10 | Industrial design | Get Messey | Shared Kitchen | 3 | Peekaboo Cam | - | - | Kitchen | 59 | 59 | Take photos of the kitchen stove when people are in the kitchen | Capture ethnographic photos for studying activities when people are cooking |
| | P33 | Industrial design | | | | | - | PIR sensor | | | | Detect human presence | To delay photo-taking for 5 minutes |
| | P74 | Automotive | | | | Sensing Door | - | PIR sensors | Kitchen Door | | | Measure the number of participants in the house. | To know how many people are in a home |
| G11 | P30 | Industrial design | Peeka Peeka Chu | Shared Kitchen & dining room | 10 | Peekaboo Cam | - | - | Kitchen | 428 | 3090 | Take photos of the dining room when participants leaves the kitchen. Take photos of the kitchen stove when participants finish using the stove and stepping away from the stove. | Capture ethnographic photos for studying cooking and dining room activities. |
| | P32 | Industrial design | | | | | Turning table | - | | | | Turn the Cam to point at the living room when participants leave the kitchen. Turn the Cam to point at the kitchen stove when participants are standing in front of the stove. | To control the camera to take photos of the in-situ activities. |
| | | | | | | | Flag | - | | | | Raise the flag to notify the Cam is preparing for photo taking; Lower the flag to notify the Cam is detecting their presence. | To notify participants that the system is detecting their presence. |
| | P63 | Electrical engineering | | | | Sensing Door | - | Ultrasonic sensor | | | | Detect human presence when they are leaving from the kitchen. | To control the camera to take photos of the in-situ activities. |
| | | | | | | Sensing box | - | Ultrasonic sensor | | | | Detect humans when they are in front of the stove. | To know when people are cooking. To control the camera to take photos of the in-situ activities. |
| | | | | | | | - | PIR sensor | | | | | |
| | P71 | Biomedical engineering | | | | Sensing Stove | - | Pressure Sensors | | | | Detect humans when they are in front of the stove. | |
| | | | | | | Kill Switch | - | - | | | | Disable the photo-taking immediately. | To give participants power to control the camera |
| G12 | P14 | Industrial design | Peekachew | Shared Fridge | 3 | Peekaboo Cam | - | - | Next to the fridge | 1099 | 162245 | Take photos when the fridge door is opened. | Capture ethnographic photos for studying human behaviours when opening the fridge door. |
| | P27 | Industrial design | | | | Sensing Fridge | - | Lighting sensors | Inside Fridge | | | Detect when the fridge is opened | To know when the fridge is opened. |
| | P35 | Industrial design | | | | Slider | - | - | Next to the fridge | | | Turn on and off the photo-taking function. | To give participants power to control the camera. |
| G13 | P46 | Industrial design | Who's Cooking | Shared Kitchen | 4 | Peekaboo Cam | - | - | Kitchen | 686 | 2120 | Take photos of the stove when the microwave turns off, the fridge has been opened, the stove is being used and the coaster is being used. | Capture ethnographic photos for studying activities when people are cooking. |
| | P48 | Software sciences | | | | | Lens Blocker | - | | | | Block the unauthorised area to be captured by the Cam | To prevent taking photos of an area that is not allowed to be captured. |
| | P55 | Biomedical engineering | | | | Microwave Cooking Timer | - | Sound sensors | | | | Detect a beep sound (at frequency of 1960Hz) from a microwave when completing a cycle | To know when a microwave is finished cooking |
| | P57 | Psychology technology | | | | Sensing Stove | - | Flame sensors | | | | Detect a flame on the stove | To know when the stove is on. |
| | | | | | | Sensing Fridge | - | Lighting sensors | | | | Detect the light of the fridge | To know when the fridge is opened. |
| | P59 | Electrical engineering | | | | Sensing Coaster | - | Temperatur e sensors | | | | Detect temperature of the mug | To know when a finished cooking meal is placed on a mug |
| G14 | P5 | Industrial design | KitcheMe | Shared Kitchen | 8 | Peekaboo Cam | - | - | kitchen | 9 | 459485 | Take photos of the kitchen stove when the lamp, water, or extractor hood are being used. | Capture ethnographic photos for studying cooking activities. |
| | P11 | Industrial design | | | | Sensing Lamp | - | Lighting sensors | | | | Detect when the lamp is turned on. | To know when the lamp is being used |
| | P28 | Industrial design | | | | Sensing Water Boiler | - | Humidity sensors | | | | Detect when the water has boiled. | To know when the water kettle is being used. |
| | P42 | Software sciences | | | | Sensing Extractor Hood | - | Sound sensors | | | | Detect when the extractor hood is being used. | To know when people are cooking |
| | | | | | | Button | - | - | | | | Disable the photo-taking immediately. | To give participants power to control the camera. |
| G15 | P43 | Software sciences | Picture your dinner | Shared Kitchen | 3 | Peekaboo Cam | - | - | kitchen | 6 | 178 | Take photos of the food/meal on the plate | Capture ethnographic photos for studying meals of a day. |
| | P49 | Software sciences | | | | Pressure Plate | - | Pressure Sensors | | | | Detect anything put on the plate | To ask participants to share their meal by putting their meal on top of the plate. |
| | P56 | Psychology technology | | | | Sensing box | - | Distance sensors | | | | Detect human presence | To know when participants are in the kitchen. "An additional LED is attached on the module. This LED lights up each time when the module sends the TRUE value to the Peekaboo Cam. This respects the privacy norm, because we let the user know that the information is sent." |
| | P75 | Web sciences | | | | | | | | | | | | |
| | P3 | Industrial design | | | | Peekaboo Cam | - | - | | | | Take photos of the clothing patterns inside a closet. | Capture ethnographic photos for studying clothing arrangement patterns in a personal closet. |

| Groups | Group Members | | Project Name | Target Context | Total Days | Name of the Data-Enabled Objects | Additional Designed Gadget | Sensors | Deployed Location | Numbers of Data Items | | Function | Purpose |
|---|---|---|---|---|---|---|---|---|---|---|---|---|---|
| | Numbers | Background | | | | | | | | Total Photos | Total Data Points* | | |
| G16 | P8 | Industrial design | Peek-a-boy | Personal Closet | 8 | Sensing Closet Railing | - | Vibration sensors | Bedroom | 33 | 162 | Measure when something is moved or changed on the closet railing | To know when people take out or put in or move the clothings. |
| | P10 | Industrial design | | | | Closet door | - | Magnetic reed-switch | | | | Detect when the closet door is opened | To know when the closet door is opened and closed. |
| | P15 | Industrial design | | | | Kill Switch | - | - | | | | Delay photo taking for few mintues. | "To give participants time to stand in front of the camera without having the fear of being photographed in a compromising way." |
| | | | | | | Sensing washing machine | - | Vibration sensors | Laundry Room | | | Detect when the washing machine is working. | To know when the washing machine is used after the closet being used. |
| G17 | P6 | Industrial design | KAIK | Shared Kitchen | 5 | Peekaboo Cam | Flag | - | Kitchen | 57 | 338 | Take photos of the stove When raising the flag, the Peekaboo Cam is put into privacy mode, which means no pictures would be taken. | Capture ethnographic photos for studying cooking activities. |
| | P60 | Electrical engineering | | | | Sensing Stove | Privacy Button | PIR sensors | | | | Detect human presence when they are in front of the stove. When the button being pressed, it will delay photo taking for five minutes. | To trigger the Peekaboo Camera: "This module does not send the measured values to the Peekaboo, but is used to control the Peekaboo." |
| | P65 | Electrical engineering | | | | Sensing Oven | - | Temperature sensors | | | | Detect when the oven is used | To know when people is cooking |
| | P68 | Mechanical engineering | | | | Sensing Fridge | - | Lighting sensors | | | | Detect when the fridge is opened | To know when people is cooking |
| G18 | P7 | Industrial design | Tadaima | Shared Home Entrance | 10 | Peekaboo Cam | - | - | Hallway | 18 | 1440 | Take photos of the home entrance | Capture ethnographic photos for studying clothing and shoes arrangement patterns after people come home. |
| | P26 | Industrial design | | | | Interactive Key holder | - | Pressure sensor | | | | Trigger photo-taking | To detect when people are entering the home. |
| | P29 | Industrial design | | | | Sensing Shoe Rack | - | Ultrasonic sensor | | | | Detect changing shoe patterns. | To know when the shoes are moved. |
| | P41 | Software sciences | | | | Sensing Plant Pot | - | PIR sensor | | | | Detect human presence | To prevent photo taking when people are in front of the Cam. |

*A data point is recorded with data type, time stamp, and data value. Data type can be distance, sound volume, as well as a Boolean expression such as "whether a photo is taken."



\end{document}